\documentclass[parskip]{article}

\newcommand{\Deutsch}{\boolean{false}}
\newcommand{\externalize}{\boolean{true}}
\newcommand{\ListAndMake}{\boolean{false}}
\newcommand{\Publishing}{\boolean{true}}

\usepackage[sort, comma, numbers]
{natbib}

\usepackage{iftex,ifthen}
\usepackage{xparse}
\usepackage{etoolbox}

\ifdef{\Deutsch}{}{\newcommand{\Deutsch}{\boolean{false}}}
\ifthenelse{\Deutsch}{
	\usepackage[ngerman]{babel}
	\AtBeginDocument{\shorthandoff{"}}
}{
	\usepackage[english]{babel}
}

\usepackage{amsmath,amssymb,amsfonts}
\usepackage{mathrsfs}
\usepackage{mathtools}
\usepackage{nicefrac}
\usepackage{physics}
\usepackage{tensor}
\usepackage{booktabs} 
\usepackage{upgreek}
\usepackage{marvosym}
\usepackage{verbatim}
\usepackage{graphicx} \graphicspath{{Graphics/},{Grafiken/},{Utilities/Graphics/}}
\usepackage{pdfpages}
\usepackage{subfiles}
\usepackage{IEEEtrantools}

\ifPDFTeX
\usepackage[T1]{fontenc}
\usepackage[utf8]{inputenc}
\usepackage[stretch=12,shrink=14,step=2,kerning=true,final]{microtype}
\else \ifLuaTeX
\usepackage[utf8]{luainputenc}
\usepackage[stretch=12,shrink=14,step=2,final]{microtype}
\else
\errmessage{Use PdfLaTeX or LuaLaTeX to compile!}
\fi
\fi

\usepackage{anyfontsize}
\usepackage{ragged2e}

\usepackage[autostyle=true]{csquotes}
\usepackage{setspace}

\renewcommand{\arraystretch}{1.15}

\let\Right\right
\let\Left\left
\makeatletter
\def\right#1{\Right#1\@ifnextchar){\!\right}{}}
\def\left#1{\Left#1\@ifnextchar({\!\left}{}}
\makeatother

\makeatletter
\g@addto@macro\bfseries{\boldmath} 
\makeatother

\makeatletter
\DeclareMathSizes{\@xiipt}{\@xiipt}{7}{5.5}
\makeatother

\DeclareUnicodeCharacter{B1}{\ensuremath{ \pm }}
\usepackage{siunitx} 
\ifPDFTeX
\newcommand{\Euro}[1]{\ifthenelse{\equal{#1}{}}{€}{\num[round-mode=places,round-precision=2,mode=text,detect-inline-family=text,detect-weight=true]{#1}\,€}}
\else \ifLuaTeX
\newcommand{\Euro}{\textbf{PDFTeX needed}}
\fi \fi

\usepackage{xcolor}

\definecolor{silver}{RGB}{235,235,235}

\definecolor{red}{RGB}{240,0,0}
\definecolor{green}{RGB}{0,240,0}
\definecolor{blue}{RGB}{0,0,240}

\colorlet{darkred}{red!60!black}
\colorlet{darkgreen}{green!60!black}
\colorlet{darkblue}{blue!60!black}

\newcommand{\FigureNames}[1]{\ifthenelse{\externalize}{\tikzsetfigurename{#1}}{}}

\newcommand{\dotp}{\ensuremath{\boldsymbol{\cdot}}}

\newcommand{\inv}[1]{\ensuremath{{#1}^{-1}}}
\newcommand{\qqq}[1]{\qquad \quad \text{#1} \quad \qquad}

\newcommand{\un}[1]{\ensuremath{\bqty{\si{#1}}}}

\ifdef{\Set}{}{\let\Set\Bqty}

\ifdef{\Avg}{}{\NewDocumentCommand{\Avg}{ o l m }{\braces#2{\langle}{\rangle}{#3}\IfNoValueTF{#1}{}{_{\!#1}}}}
\input{arXiv/referencing}

\usepackage{pgfplots,pgfplotstable,mathtools,IEEEtrantools,physics,tensor,stmaryrd,upgreek,amsfonts,amssymb,xparse,csquotes}

\usepackage{caption}
\usepackage[labelformat=simple]{subcaption}

\makeatletter
\@ifpackageloaded{caption}{\captionsetup{labelfont={bf,color=figcol}}}{}
\makeatother

\usetikzlibrary{arrows, external, calc}
\tikzstyle{every picture}+=[line join=round,line cap=round]

\tikzset{>=latex}

\ifthenelse{\externalize}{%
	\usepgfplotslibrary{external}%
	\tikzexternalize[prefix=TikZ/]%
	\ifthenelse{\ListAndMake}{\tikzset{external/mode=list and make}}{}%
}{}

\usepgfplotslibrary{patchplots,fillbetween,groupplots,colormaps}

\usetikzlibrary{shadows,arrows.meta,calc,angles,shapes,quotes,babel,math,decorations.pathreplacing,calligraphy,matrix}

\pgfplotsset{compat = 1.17}
\pgfplotsset{grid=major, enlarge y limits=0.05, enlarge x limits=false, every axis/.append style={line join=round}}

\pgfplotsset{width=0.75\textwidth, height=0.4\textwidth}

\definecolor{SeqCol1}{RGB}{166,206,227} 
\definecolor{SeqCol2}{RGB}{31,120,180 } 
\definecolor{SeqCol3}{RGB}{178,223,138} 
\definecolor{SeqCol4}{RGB}{51,160,44  } 
\definecolor{SeqCol5}{RGB}{251,154,153} 
\definecolor{SeqCol6}{RGB}{227,26,28  } 
\definecolor{SeqCol7}{RGB}{253,191,111} 
\definecolor{SeqCol8}{RGB}{255,127,0  } 
\definecolor{SeqCol9}{RGB}{202,178,214} 
\definecolor{SeqCol10}{RGB}{106,61,154} 

\definecolor{PastellSeqCol1}{RGB}{141,211,199}
\definecolor{PastellSeqCol2}{RGB}{255,255,179}
\definecolor{PastellSeqCol3}{RGB}{190,186,218}
\definecolor{PastellSeqCol4}{RGB}{251,128,114}
\definecolor{PastellSeqCol5}{RGB}{128,177,211}
\definecolor{PastellSeqCol6}{RGB}{253,180,98 }
\definecolor{PastellSeqCol7}{RGB}{179,222,105}
\definecolor{PastellSeqCol8}{RGB}{252,205,229}
\definecolor{PastellSeqCol9}{RGB}{217,217,217}
\definecolor{PastellSeqCol10}{RGB}{188,128,189}

\definecolor{mycol6}{HTML}{FFBB00} 

\colorlet{mycol1}{SeqCol6}
\colorlet{mycol2}{SeqCol2}
\colorlet{mycol3}{SeqCol4}
\colorlet{mycol4}{SeqCol8}
\colorlet{mycol5}{SeqCol10}

\colorlet{fitcol}{mycol4} 
\colorlet{plotgrey}{black!75!white} 

\colorlet{mycol1light}{SeqCol5}
\colorlet{mycol2light}{SeqCol1}
\colorlet{mycol3light}{SeqCol3}
\colorlet{mycol4light}{SeqCol7}
\colorlet{mycol5light}{SeqCol9}

\pgfplotscreateplotcyclelist{mycolors}{%
	mycol1, no marks, thick\\%
	mycol2, no marks, thick\\%
	mycol3, no marks, thick\\%
	mycol4, no marks, thick\\%
	mycol5, no marks, thick\\%
	mycol6, no marks, thick\\%
}
\pgfplotsset{every axis/.append style={cycle list name=mycolors}}

\newcommand{\InfoGeo}{\href{https://github.com/RafaelArutjunjan/InformationGeometry.jl}{\texttt{InformationGeometry.jl}}}

\newcommand{\M}{\ensuremath{\mathcal{M}}}
\newcommand{\N}{\ensuremath{\mathcal{N}}}
\newcommand{\R}{\ensuremath{\mathbb{R}}}
\renewcommand{\P}{\ensuremath{\mathbb{P}}}

\newcommand{\X}{\ensuremath{\mathcal{X}}}
\newcommand{\Y}{\ensuremath{\mathcal{Y}}}

\NewDocumentCommand{\E}{ o l m }{\mathbb{E}\IfNoValueTF{#1}{}{_{#1}}\braces#2{\lparen}{\rparen}{#3}}
\NewDocumentCommand{\Var}{ o l m }{\mathrm{Var}\IfNoValueTF{#1}{}{_{#1}}\braces#2{\lparen}{\rparen}{#3}}

\newcommand{\qqc}{\qc\qquad}
\newcommand{\openslot}{\ensuremath{\,\cdot\,}}
\newcommand{\data}{\mathrm{data}}
\newcommand{\model}{\mathrm{model}}

\newcommand{\VF}[1]{\ensuremath{\Gamma\pqty{T #1}}}
\newcommand{\CoVF}[1]{\ensuremath{\Gamma\pqty{T^* #1}}}
\newcommand{\Lie}[1]{\ensuremath{\mathcal{L}_{#1}\,}}
\newcommand{\Cov}[1]{\ensuremath{\nabla_{\!#1}\,}}
\newcommand{\Smooth}[1]{\ensuremath{C^\infty\pqty{#1}}}

\newcommand{\Complexity}[1]{\ensuremath{\mathcal{O}\pqty{#1}}}

\DeclareDocumentCommand{\cond}{ s m g }
{\! 
	\IfBooleanTF{#1}
	{ 
		\IfNoValueTF{#3}
		{\vphantom{#2}\left(\smash{#2}\,\middle\vert\,\smash{#2}\right)}
		{\vphantom{#2#3}\left(\smash{#2}\,\middle\vert\,\smash{#3}\right)}
	}
	{ 
		\IfNoValueTF{#3}
		{\left({#2}\,\middle\vert\,{#2}\right)}
		{\left({#2}\,\middle\vert\,{#3}\right)}
	}
}

\DeclareDocumentCommand{\Setcond}{ s m g }
{\! 
	\IfBooleanTF{#1}
	{ 
		\IfNoValueTF{#3}
		{\vphantom{#2}\left\lbrace\smash{#2}\,\,\middle\vert\,\,\smash{#2}\right\rbrace}
		{\vphantom{#2#3}\left\lbrace\smash{#2}\,\,\middle\vert\,\,\smash{#3}\right\rbrace}
	}
	{ 
		\IfNoValueTF{#3}
		{\left\lbrace{#2}\,\,\middle\vert\,\,{#2}\right\rbrace}
		{\left\lbrace{#2}\,\,\middle\vert\,\,{#3}\right\rbrace}
	}
}

\NewDocumentCommand{\Div}{ o m }{\ensuremath{\IfNoValueTF{#1}{D}{D_{#1}}\bqty{#2}}}

\newcommand{\JuliaPkgBase}[1]{{\textcolor{black}{\texttt{#1.jl}}}}
\NewDocumentCommand{\jlpkg}{ o m }{\IfNoValueTF{#1}{\JuliaPkgBase{#2}}{\defcitealias{#1}{\JuliaPkgBase{#2}}\citetalias{#1}}}

\let\oldhat\hat
\renewcommand{\hat}[1]{\smash{\oldhat{#1}}\vphantom{'}}

\DeclareMathOperator{\preim}{\ensuremath{\mathrm{preim}}}

\tikzstyle{mpoint}=[fill=fitcol,inner sep=1pt,circle]

\usepackage{arXiv/arxiv}	\graphicspath{{arXiv/}}

\usepackage{doi}

\NewDocumentCommand{\PublishGraphic}{ o m m }{%
	\ifthenelse{\Publishing}{
		\IfNoValueTF{#1}{%
			\includegraphics{#2.pdf}
		}{%
			\includegraphics{#1/#2.pdf}
		}
	}{
		\ifthenelse{\externalize}{\tikzsetnextfilename{#2}}{}%
		#3%
}}

\title{Constructing Exact Confidence Regions on\\ Parameter Manifolds of Non-Linear Models}

\NewDocumentCommand{\MailTo}{O{} m}{%
	\ifthenelse{\equal{#1}{}}{\href{mailto:#2}{\Letter~\texttt{#2}}}{\href{mailto:#2?subject=#1}{\Letter~\texttt{#2}}}
}

\author{ \href{https://orcid.org/0000-0002-6658-4120}{\includegraphics[height=0.75em]{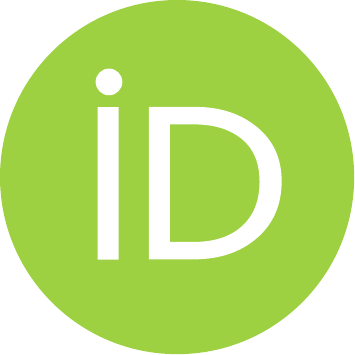}\hspace{1mm}Rafael Arutjunjan} 
	\\
	Institute of Physics\\
	University of Freiburg\\
	Hermann-Herder-Str.\ 3, 79104 Freiburg, Germany\\
	\MailTo[Re: Exact Confidence Regions]{rafael.arutjunjan@fdm.uni-freiburg.de} \\
	\And
	\href{https://orcid.org/0000-0002-9453-5772}{\includegraphics[height=0.75em]{orcid.pdf}%
	\hspace{1mm}Björn Malte Schäfer} \\
	Zentrum für Astronomie der Universität Heidelberg\\
	Astronomisches Rechen-Institut\\
	Philosophenweg 12, 69120 Heidelberg\\
	\MailTo{bjoern.malte.schaefer@uni-heidelberg.de} \\
	\And
	\href{https://orcid.org/0000-0002-8796-5766}{\includegraphics[height=0.75em]{orcid.pdf}\hspace{1mm}Clemens Kreutz} \\
	Institute of Medical Biometry and Statistics\\
	Faculty of Medicine and Medical Center\\
	University of Freiburg\\
	\MailTo{ckreutz@imbi.uni-freiburg.de} \\
}

\date{}

\renewcommand{\headeright}{}
\renewcommand{\undertitle}{}
\renewcommand{\shorttitle}{Constructing Exact Confidence Regions on Parameter Manifolds of Non-Linear Models}

\hypersetup{
	pdftitle={Constructing Exact Confidence Regions on Parameter Manifolds of Non-Linear Models},
	pdfsubject={physics.data-an, astro-ph.CO, math.ST}, 
	pdfauthor={Rafael Arutjunjan, Bjoern Malte Schaefer, Clemens Kreutz},
	pdfkeywords={Confidence Regions, Confidence Bands, Information Geometry, Fisher Metric, Parameter Inference},
}

\begin{document}
	\maketitle

	\begin{abstract}
	Using the mathematical framework of information geometry, we introduce a novel method which allows one to efficiently determine the exact shape of simultaneous confidence regions for non-linearly parametrised models. Furthermore, we 
	show how pointwise confidence bands around the model predictions can be constructed from detailed knowledge of the exact confidence region with little additional computational effort. 
	We exemplify our methods using inference problems in cosmology and epidemic modelling. An open source implementation of the developed schemes is publicly available via the \InfoGeo\ package for the \href{https://julialang.org}{Julia programming language}.
	\end{abstract}



\FigureNames{Introduction}

\section{Introduction}

The goal of parameter inference is not only to find optimal parameter values such that a given model best describes observational data, but also to subsequently use this model to make predictions for the outcomes of future experiments. However, since any observation in the real world features stochastic noise, providing precise quantifications of the uncertainties associated with the parameters is a vital part of the inference process, such as to not render the predictions of a model ultimately meaningless. This quantification of parameter uncertainties is typically achieved by establishing confidence regions around the parameter configuration corresponding to the best fit. The uncertainties in the model predictions can subsequently be calculated from the parameter uncertainties.

In many applications, researchers rely on approximations of confidence regions e.g.\ by using the Cramér--Rao inequality, which states that in the large sample limit, a lower bound for the covariance matrix associated with the parameters is given by the inverse of the Fisher information matrix as evaluated at the best fit \cite{CramerRaoLowerBound}. 
For models which depend linearly on their parameters and observations with Gaussian (i.e.\ normal) noise distributions, it is straightforward to show that the confidence regions are always given by perfect ellipsoids centered on the maximum likelihood estimate (MLE) in the parameter space. Since any $n$-dimensional ellipsoid is related to the unit $n$-sphere via a unique affine transformation, the size and shape of ellipsoidal confidence regions relative to the MLE can be fully encoded using a symmetric positive-definite matrix, that is, a covariance matrix for the estimated parameters. In contrast, confidence regions associated with models which depend non-linearly on their parameters are no longer of ellipsoidal shape, but are non-linearly distorted. The magnitude of this distortion depends on both the given parametrisation of the model and also on the quality and amount of available experimental data. 

Since a matrix is no longer sufficient for capturing the distorted shapes of confidence regions for non-linearly parametrised models, it is clear that the Cramér--Rao lower bound cannot provide an accurate quantification of the true simultaneous parameter uncertainties. 
While many questions relating to maximum likelihood estimation, systematic model reduction and optimal experimental design have been discussed by numerous publications in the past, the topics of parameter uncertainty and confidence regions remain incompletely addressed by the available information-geometric literature.



Given that for non-linearly parametrised models the shapes of confidence regions often strongly vary depending on their associated confidence level (see e.g.\ \cref{fig:Reparametrisations} in \cref{sec:Survey}), linearised approximations of the parameter uncertainties via a constant covariance matrix can be misrepresentative of the underlying sensitivity of the model with respect to changes in the parameter values.
Thus, in applications where a nuanced 
understanding of the model parameters and their interdependence is required, a more elaborate investigation must be conducted to determine the exact extents of confidence regions. In this context, \enquote{exact} refers to the fact that the confidence regions are not only simultaneous, meaning that the interactions between the various parameters are taken into account, but that no simplifying assumptions are made about the shapes of the confidence regions.


Current state of the art methods for constructing exact simultaneous confidence regions rely on evaluating the likelihood for a multitude of parameter configurations $\theta \in \M$ either on a grid or stochastically \cite{Leibundgut}. For this reason, accurate constructions of exact confidence regions have typically been considered to be computationally infeasible, particularly for complex models and large datasets.

In this article, we propose an efficient numerical scheme for the construction of confidence boundaries for non-linearly parametrised models. In essence, this scheme converts the problem of locating a confidence boundary associated with some confidence level $q \in (0,1) \subset \R$ into numerically solving a system of ordinary differential equations (ODEs). Its improved performance compared with previously established methods essentially results from the fact that it does not require sampling of the likelihood over large volumes in the parameter space either on a grid or stochastically. We also provide a proof which highlights the structural identifiability of the model as the only necessary criterion for the applicability of the presented method. Moreover, we show how knowledge of the exact confidence boundaries can be used to obtain confidence bands around the model predictions with minimal additional computational effort.


Also, while the distribution underlying the uncertainties in the observed data is required to be unimodal, 
the proposed method is agnostic with respect to the precise shape of the distribution. In other words, the proposed method is not only applicable for observed data with Gaussian uncertainties but also other distributions such as student's $t$-distributions or even asymmetric distributions.

An open source implementation of the presented methods is publicly available in the form of the \InfoGeo\ package for the \href{https://julialang.org}{Julia programming language} \cite{InformationGeometryJL}. The discussions in this article as well as some figures closely follow \cite{GeometricParameterInference}.


%
%
%
%
%

\section{Methodology}

In this section, we briefly summarise relevant definitions and terminology from the subject of parameter inference. 
For a technical review of core concepts of differential geometry such as coordinate charts, Riemannian metrics, geodesics and curvature, we refer to standard literature on the subject such as \cite{Isham, Lee, Hall, MurrayRice, AmariBook1}.

\subsection{Information Divergences and the Fisher Metric} \label{sec:DivergencesAndFisherMetric}


Information divergences are positive-definite functionals which are used to compare probability distributions with common support and which quantify a notion of separation or dissimilarity \cite{GeodesicAcceleration, Tegmark, AmariBook1, GeometricParameterInference}. The canonical example is given by the Kullback--Leibler divergence $D_\text{KL}$ defined by
\begin{equation} \label{eqn:DKL}
	D_\text{KL}[p, q] \coloneqq	\int \! \dd{y} p(y) \, \ln(\frac{p(y)}{q(y)})
\end{equation}
which can be interpreted as quantifying the relative increase in Shannon entropy (i.e.\ loss of information) from approximating a probability distribution $p$ via another distribution $q$. Although the Kullback--Leibler divergence has many desirable properties, it should be noted that it does not induce a proper notion of distance between distributions since it is neither symmetric with respect to its arguments nor satisfies a triangle inequality \cite{AmariBook1}.

When restricting attention to pairs of distributions within a single family of probability distributions that can be parametrised using a finite number of parameters $\theta = (\theta_1,...,\theta_n) \in \M \subseteq \R^n$ with respect to which they are differentiable, the so-called Fisher information matrix is defined as the Hessian of the Kullback--Leibler divergence via
\begin{equation} \label{eqn:FisherMetricFromDKL}
	\tensor{g}{_a _b}(\theta) \coloneqq \bqty{\pdv{}{\psi^a \,}{\psi^b} \, D_\text{KL}\bqty\big{ p(y;\theta), p(y;\psi)}}_{\psi = \theta} = ... = -\E[p(y;\theta)\!]{\pdv{\ln(p)}{\theta^a\,}{\theta^b}}. 
\end{equation}
By expanding the Kullback--Leibler divergence in a Taylor series with respect to the parameters, one finds that the zeroth and first order terms vanish, wherefore the Fisher information matrix fully encodes an infinitesimal approximation to the Kullback--Leibler divergence as it is the first non-vanishing coefficient in this expansion.

Since the Fisher information matrix provides a symmetric, positive-definite bilinear form at every point $\theta \in \M$ and exhibits the transformation behaviour of a $(0,2)$-tensor field, it can be seen as constituting a Riemannian metric on \M\ \cite{Lee}. Given its direct relationship to the Kullback--Leibler divergence, it is also evident that this is a very special choice of metric. 
Furthermore, it was first proven by \v{C}encov that the Fisher metric is in fact the unique Riemannian metric (up to rescaling) which is invariant under a class of probabilistically meaningful embeddings known as Markov morphisms \cite{Cencov,LebanonExtended}.


In the applied context of the following discussions, the parametrised families of probability distributions which induce the Fisher metric are likelihoods which compare the output of some mathematical model against observed data. Denoting the domains of the independent and dependent variables by $\X$ and $\Y$ respectively, the model constitutes a map $y_\model : \X \cross \M \longrightarrow \Y$. For observations with Gaussian uncertainties, the log-likelihood $\ell = \mathrm{ln} \circ L$ is given by
\begin{equation} \label{eqn:GaussianLikelihood}
	\ell\cond{\mathrm{data}}{\theta} = -\frac{N}{2} \ln(2\pi) -\frac{1}{2}(\ln\circ\det)\pqty{\Sigma} \, -\frac{1}{2} \, \tensor{\pqty{y_\data - h(\theta) \vphantom{)^2}}}{^a} \, \tensor{(\inv{\Sigma})}{_a _b} \, \tensor{\pqty{y_\data - h(\theta) \vphantom{)^2}}}{^b}
\end{equation}
where $y_\data \coloneqq (y_1, ..., y_N) \in \Y^N$ denotes the vector of concatenated observations with $\Sigma$ the associated covariance matrix, $h : \M \longrightarrow \Y^N$ is the embedding map defined by $h(\theta) \coloneqq \pqty\big{y_\model(x_1;\theta),...,y_\model(x_N;\theta)}$ and the Einstein summation convention is employed. For the special case of likelihoods which are of the form exemplified in \cref{eqn:GaussianLikelihood}, a practical expression for the Fisher metric can be derived 
as
\begin{equation} \label{eqn:GaussianFisherMetric}
	\tensor{g}{_a _b}(\theta) = -\int_{\Y^N} \! \dd{y_\data} L\cond{y_\data}{\theta} \, \pdv{\ell}{\theta^a\,}{\theta^b} = \pdv{h^i}{\theta^a} \, (\Sigma^{-1})_{ij} \, \pdv{h^j}{\theta^b}
\end{equation}
where a constant covariance matrix $\Sigma$ was assumed for the observations, i.e.\ $\pdv*{\Sigma}{\theta} = 0$. Furthermore, by recognising that for Gaussian observations, $g_{\Y^N} = \inv{\Sigma}$ constitutes an inner product on the data space $\Y^N$, one finds that the expression derived for the Fisher metric in \cref{eqn:GaussianFisherMetric} coincides exactly with the pull-back metric $h^* g_{\Y^N}$ under the map $h$. For this reason, the term \enquote{embedding map} is justified whenever the model map $y_\model$ from which $h$ is built is injective with respect to the parameters $\theta \in \M$. 


\subsection{Defining Confidence Regions} \label{sec:DefiningConfidenceRegions}

Various alternative definitions of simultaneous confidence regions have been proposed in the past, each of which comes with advantages and drawbacks. The most widely used definitions of confidence regions are based on hypothesis tests such as the likelihood ratio test or the $F$-test \cite{SeberConfidence, CaseStudy}.


A confidence region of level $q \in (0,1) \subset \R$ is a set of parameter configurations which is judged to contain the \enquote{true} parameter configuration, i.e.\ the parameter configuration which most likely underlies the observations, with probability $q$. That is, if the experiment producing the observations were to be repeated many times, the \enquote{true} parameter configuration is contained in a fraction $q$ of the confidence region realisations of confidence level $q$. In addition, we define a confidence region as not just any set which contains the \enquote{true} parameter configuration with probability $q$: instead, it should only contain the parameter configurations which give the best possible descriptions of the data. That is, there should be no parameter configuration outside the confidence region that describes the observed data as well or better than any parameter configuration inside the confidence region (e.g.\ has a higher likelihood). By exploiting detailed knowledge about its parent distribution, one can then determine suitable threshold values for a test statistic below or above which the test is said to reject a hypothesis with confidence level $q$.


\subsubsection{Confidence Regions Based on the Likelihood Ratio Test} \label{sec:LikelihoodRatioTestRegions}

The definition of confidence regions via the likelihood ratio test utilises Wilks' theorem \cite{Wilks}, which states that the log-likelihood difference is asymptotically distributed according to $\chi^2_k$ with $k$ the degrees of freedom, i.e.\ the number of components in which $\theta$ differs from $\theta_\text{MLE}$. More precisely, Wilks showed that $2\pqty\big{\ell(\theta_\text{MLE}) - \ell(\theta)} \sim \chi^2_k$ in the large sample limit, i.e.\ as the number of data points $N \longrightarrow \infty$. Thus, a confidence region of level $q \in (0,1) \subset \R$ on the parameter manifold \M\ may be defined as
\begin{equation} \label{eqn:ConfidenceRegionDefinition}
	\mathcal{C}_q \coloneqq \,\Setcond{\theta \in \M}{ 2\pqty\big{\ell\pqty{\theta_\text{MLE}} - \ell(\theta)} \leq F_k^{-1}(q)} = \,\Setcond{\theta \in \M}{F_k\pqty\Big{2\bqty\big{\ell(\theta_\text{MLE}) - \ell(\theta)}} \leq q}
\end{equation}
where $F_k^{-1}$ denotes the inverse cumulative distribution function of the $\chi^2_k$ distribution in this context (i.e.\ its quantile function), with $k$ the degrees of freedom and $\ell$ is the log-likelihood. 

The popularity of this approach stems at least in part from the Neyman--Pearson lemma, which guarantees that the likelihood ratio test is the most powerful test when comparing simple hypotheses \cite{NeymanPearson}. In addition, the likelihood ratio test is parametrisation-invariant and applicable in most practical settings. The boundary of a confidence region $\partial \mathcal{C}_q$ is then given by
\begin{equation} \label{eqn:ConfidenceBoundaryDefinition}
	\partial \mathcal{C}_q = \,\Setcond{\theta \in \M}{ 2\pqty\big{\ell\pqty{\theta_\text{MLE}} - \ell(\theta)} = F_k^{-1}(q)}.
\end{equation}
Since $\ell\pqty{\theta_\text{MLE}}$ is a constant, it is straightforward to see that the confidence boundaries correspond to the level sets of the likelihood function. Therefore, while the large sample limit assumed in Wilks' theorem might not always apply, this only impacts which confidence level $q$ is associated to a given level set of the likelihood, but not the shapes of confidence regions themselves. Consequently, one can try to conservatively overestimate the size of a confidence region to compensate for small sample sizes without affecting the model sensitivity information encoded in the shape of the confidence region. 
Although we focus on likelihood-based confidence regions in this work, the integral manifold method outlined in \cref{sec:IsoLikelihoodSurfaces} can be applied whenever the confidence boundary is defined in terms of the level sets of a function whose Hessian with respect to the parameters is non-singular on the domain of interest.


Depending on the model function and available data, the likelihood may be multimodal, meaning that it can have more than one local maximum which can potentially result in topologically disconnected confidence regions. Although different authors disagree on whether topologically disconnected confidence regions are reasonable, it is often sensible to require that any point $\theta \in \mathcal{C}_q$ be (path-)connected to $\theta_\text{MLE}$ on top of the definition in \cref{eqn:ConfidenceRegionDefinition}. 
Moreover, the existence of multiple local maxima in the likelihood, which ultimately leads to disconnected confidence regions, can often be traced back to a global structural non-identifiability of some kind in the model parametrisation (see \cref{sec:ParameterIdentifiability}). Such global structural non-identifiabilities can in principle be remedied by suitably restricting the parameter domain although this is not always straightforward in practice \cite{VillaverdeLieSymmetries}.

%
%
%
%
%

\subsection{Structural and Practical Parameter Identifiability} \label{sec:ParameterIdentifiability}

We briefly summarise some widely-used terminology surrounding the topic of parameter identifiability which is discussed in more depth for instance in \cite{GeodesicAcceleration, Timmer, Sloppiness, SloppyVSIdentifiable}. 

A model is said to be locally structurally identifiable at a point $\theta \in \M$ if there exists a non-empty neighbourhood $U$ around $\theta \in \M$ where no other parameter configuration $\psi \in U$ results in the same model prediction as $\theta$. It has been demonstrated \cite{Rothenberg} that 
\begin{equation}
	\mathrm{det}\pqty\big{g(\theta)} \neq 0 \qquad \Longleftrightarrow \qquad \text{model is locally structurally identifiable at } \theta \in \M
\end{equation}
with $g$ the Fisher metric defined in \cref{eqn:FisherMetricFromDKL}. Therefore, the non-vanishing determinant of the Fisher metric provides a practical and coordinate-invariant criterion which encodes whether a model is locally injective with respect to its parameters. 
By investigating Lie symmetries of a model with respect to vector fields on the parameter manifold \M\ in more detail, it is possible to systematically construct symmetry-breaking transformations that allow for model reformulations by which non-identifiable models can be made locally structurally identifiable \cite{VillaverdeLieSymmetries}. 
Moreover, if a model is injective on the entire domain, it is said to be globally structurally identifiable. However, verifying the global injectivity of a model is often a laborious process and may be infeasible in practice for models with high complexity, given that there is no convenient criterion which can be checked for this.

Compared to the concept of structural identifiability, it is more difficult to come up with a quantitative definition of practical identifiability. Generally, it should encapsulate the phenomenon that some parameters of the model are not suitably constrained by the available data to make definitive statements about their values for all confidence levels $q$. That is, their one-dimensional confidence intervals of level $q$ are either not bounded from below, above or both.

When evaluating the log-likelihood along the radial path of slowest descent starting at the MLE, its value is sometimes bounded from below along this path. As a result, there is some confidence level $q$ for which the log-likelihood values on the path of slowest descent are too close to the value at the MLE for a difference larger than $\frac{1}{2} \, F_k^{-1}(q)$ to be attained. Thus, the threshold which defines the confidence boundary $\partial \mathcal{C}_q$ is not crossed along this radial path of slowest descent, which means that the confidence region is unbounded in this direction. On the other hand, for models which are structurally identifiable at the MLE, the resulting negative-definiteness of the Hessian of the log-likelihood ensures that there exists some $q > 0$ such that the associated confidence region $\mathcal{C}_q$ is bounded.

By this definition of practical identifiability, it is clear that local structural non-identifiability directly implies practical non-identifiability, due to the existence of a direction along which the likelihood is constant. Practical non-identifiabilities are particularly straightforward to detect via the so-called profile likelihood method \cite{Timmer}.

\section{Results}

As argued before, detailed knowledge of the exact confidence regions provides richer insight into the interdependence of the various model parameters for non-linearly parametrised models in contrast to approximations such as the Cramér--Rao lower bound. In this section, we describe an efficient scheme for locating exact confidence boundaries using established concepts of information geometry.

\subsection{Geometric Construction of Iso-Likelihood Surfaces} \label{sec:IsoLikelihoodSurfaces}

The method outlined within this section demonstrates how the definition of confidence regions based on the level sets of some function $f \in \Smooth{\M}$ can be exploited to find the exact boundaries of said confidence regions in a numerically efficient way. The general idea is to try to systematically construct complete vector fields which are tangential to the level sets of $f$ such that their integral curves or surfaces can be used to recover the entire level set.

This turns the problem of finding the boundary of a confidence region into a system of ordinary differential equations which can then be solved using numerical methods. The desired confidence level of the boundary is specified by supplying a point which is already known to lie on said boundary as an initial condition for the system of ODEs. 
This represents a significant reduction in computational effort, since the likelihood ratio test only needs to be evaluated on a one-dimensional line emanating from the maximum likelihood configuration $\theta_\text{MLE} \in \M$ to find such a point. Although this method was developed with the application of constructing confidence boundaries in mind, it can be used to parametrise the level sets of any smooth function which satisfies the requirements discussed in \secref[sec:LieAlgebraClosure]{appendix \ref*{sec:LieAlgebraClosure}}.

Given a scalar function $f \in \Smooth{\M}$, its gradient is calculated using the exterior derivative, resulting in a covector field $\dd f \in \CoVF{\M}$. Given such a covector field, one can try to find a vector field $X \in \VF{\M}$ such that in a chart $(U,\theta)$
\begin{equation} \label{eqn:ScoreAnnihilation}
	(\dd f)(X) = X^j \, \pdv{f}{\theta^j} \overset{!}{=} 0  \qquad \text{everywhere.}
\end{equation}
In other words, the vector field $X$ is annihilated by the gradient of $f$ at every point. 
One might ponder the question of whether there are alternative principled ways of constructing vector fields which are tangential to the level sets of $f$, for example whether the construction should somehow account for geometric properties of \M\ like curvature using the covariant derivative $\Cov{X}\!$. However, since both the covariant derivative $\Cov{X}\!$ and also the Lie derivative $\Lie{X}\!$ of a smooth function with respect to a vector field $X \in \VF{\M}$ by definition reduce to the same behaviour as the vector field $X$ acting on the function, one ends up with exactly the same criterion: 
\begin{equation}
	\Cov{X} f = \Lie{X} f  = (\dd f)(X) = X f.
\end{equation}
Intuitively, every one of these formulations aims to find a vector field along which the function $f$ does not change in value. Disregarding the trivial vector field $X=0$, a reasonable strategy for finding a general solution to \cref{eqn:ScoreAnnihilation} is to choose the components of $X$ as
\begin{equation} \label{eqn:OrthVF}
	X^j = \alpha^j \, \prod_{i \neq j} \pdv{f}{\theta^i} \qqq{for} \alpha^j \in \R : \sum_j \alpha^j = 0 \qq{and} j=1,...,\dim \M.
\end{equation}
Inserting this form of $X$ into \cref{eqn:ScoreAnnihilation}, one finds
\begin{equation} \label{eqn:AlphaCondition}
	X^j \, \pdv{f}{\theta^j} = \pqty{\sum_{j=1}^{\dim\M} \alpha^j} \underbrace{\prod_{i=1}^{\dim\M} \pdv{f}{\theta^i}}_{\eqqcolon B} \overset{!}{=} 0
\end{equation}
which, given that the product amounting to $B$ is non-zero for locally structurally identifiable models away from the MLE, vanishes exactly if $\sum_j \alpha^j = 0$. Moreover, one can see that for functions $f$ which are $k$ times differentiable, the resulting vector field $X$ will be $k-1$ times differentiable. That is to say, $X$ is smooth if $f$ is smooth.

The condition $\sum_j \alpha^j \overset{!}{=} 0$ can be geometrically interpreted as a $(\dim\M -1)$-dimensional hyperplane $\mathcal{H}$ in the real vector space $\R^{\dim{\M}}$ equipped with the standard inner product:
\begin{equation}
	\mathcal{H} \coloneqq \,\Setcond{\vec{\alpha} \in \R^{\dim\M}}{\textstyle{\sum_j} \, \alpha^j = 0} = \,\Setcond{\vec{\alpha} \in \R^{\dim\M}}{\vec{n} = (1,...,1)^\top,~ \vec{\alpha} \dotp \vec{n} = 0} = \pqty\big{\mathrm{span}\Bqty{\vec{n}}}^\bot.
\end{equation}
This shows that any $\vec{\alpha}$ which is orthogonal to $\vec{n} = (1,...,1)^\top$ with respect to the standard inner product on $\R^{\dim\M}$ provides a solution to \cref{eqn:AlphaCondition}. Since by definition a vector space is $n$-dimensional if and only if it admits a set of $n$ linearly independent basis vectors, it is clear that the hyperplane $\mathcal{H}$ must contain $\dim{\M}-1$ vectors which are mutually orthogonal, as well as orthogonal to $\vec{n}$ wherefore $\dim{\mathcal{H}} = \dim{\M}-1$.

Since linear independence is preserved under vector space isomorphisms, the frame obtained by mapping a basis of $\R^{\dim\M}$ under $K_p$ is guaranteed to span $T_p\M$. In particular, any vector field $X \in K(\mathcal{H}) \subset T\M$ generated from elements of $\mathcal{H}$ satisfies the desired condition \cref{eqn:ScoreAnnihilation}. Specifically in the case where the log-likelihood function $f=\ell$ is considered, one can read off from \cref{eqn:OrthVF} that the vector space isomorphism $K_p : \R^{\dim\M} \longrightarrow T_p\M$ must be given by
\begin{equation} \label{eqn:VSisomorphism}
	\bqty{K_p(\vec{\alpha})}^{i} = \tensor{M}{^i _j} \, \alpha^j = \bqty{\prod_{k=1}^{n} \pdv{\ell}{\theta^k}} \, \mathrm{diag}\pqty{\! \pqty{\pdv{\ell}{\theta^1}}^{\!-1}, ..., \pqty{\pdv{\ell}{\theta^n}}^{\!-1} }^{\!\!i}_{\,j} \,\, \alpha^j
\end{equation}
where $n = \dim\M$. As mentioned previously, for the case of structurally identifiable models and unimodal error distributions, the components of the gradient of the log-likelihood vanish only at the MLE. 
The integral curves of $X$ will then trace out level sets $I_c$ of the log-likelihood $\ell$ defined by
\begin{equation}
	I_c \coloneqq \preim_\ell(c) = \,\Setcond{\theta \in \M}{\ell(\theta)=c}
\end{equation}
given an initial condition in the form of a starting point which already lies on the desired level set. The defining equation for an integral curve $\gamma$ to a vector field $X$ is given by
\begin{equation} \label{eqn:IntegralCurve}
	X_{\gamma(t)} \overset{!}{=} \dot{\gamma}(t)
\end{equation}
which enforces that the tangent vectors $\dot{\gamma}$ to the curve $\gamma$ coincide with the vector field $X$ at every point through which the curve passes. This condition translates to a set of ordinary differential equations that is guaranteed to have a unique solution (at least locally) by virtue of the Picard--Lindelöf theorem, given appropriate initial conditions. More generally, the existence of integral surfaces or integral manifolds is characterised by the Frobenius theorem \cite{Lee}, whose requirement that the set of generating vector fields should span a closed Lie algebra is trivially fulfilled in the one-dimensional case.

A proof that the set of all vector fields constructed according to \cref{eqn:AlphaCondition} forms a closed Lie subalgebra of $\Gamma(T\M)$ and therefore the integral manifolds generated by such vector fields foliate \M\ is given in \secref[sec:LieAlgebraClosure]{appendix \ref*{sec:LieAlgebraClosure}}. The proof also highlights the structural identifiability of the model on the closure of the desired confidence region $\overline{\mathcal{C}_q}$ as well as the twice differentiability of $\ell$ with respect to the parameters $\theta \in \M$ as the only necessary requirements for the proposed scheme.

%
%
%

\subsection{Confidence Bands} \label{sec:ConfidenceBands}

Since the ultimate goal of assessing parameter uncertainty is to determine the uncertainty in the model predictions, we show how this can be achieved efficiently, given knowledge of the exact confidence regions associated with a maximum likelihood estimate. 
In many publications (see e.g.\ \cite{ConfidenceBands}), one finds a definition of confidence bands along the following lines: Two functions $l(x)$ and $u(x)$ constitute the boundary of a pointwise confidence band of confidence level $q$ around a model $y_\model(x;\theta_\text{MLE})$ if
\begin{equation} \label{eqn:PWConfBands1DDef}
	\forall x \in \X: \qquad \P\bqty\big{l(x) \leq y_\model(x;\theta_\text{MLE}) \leq u(x)} = q.
\end{equation}
That is, at each $x\in \X$, the interval $[l(x),u(x)] \subseteq \Y \subseteq \R$ separately provides a confidence interval around the prediction $y_\model(x;\theta_\text{MLE})$ of the model function. Importantly, pointwise confidence bands are not to be confused with simultaneous confidence bands which, in contrast, are defined as
\begin{equation}
	\P\bqty\big{\forall x \in \X : \quad l(x) \leq y_\model(x;\theta_\text{MLE}) \leq u(x)} = q
\end{equation}
which differs only subtly from the definition of pointwise confidence bands in its placement of the \enquote{$\forall x \in \X$} qualification.


Apart from the fact that the definition of pointwise confidence bands in \cref{eqn:PWConfBands1DDef} is only applicable for one-dimensional dependent variables, i.e.\ when $\dim\Y = 1$, it also does not provide a practical recipe for calculating said confidence bands. Arguably, a more practical definition of a pointwise confidence band of level $q$ is given by
\begin{equation} \label{eqn:PWConfBandsDef}
	\mathcal{B}_q(x) \coloneqq y_\model(x;\mathcal{C}_q) = \,\Setcond{y_\model(x;\theta)\in\Y}{\theta \in \mathcal{C}_q}
\end{equation}
which generalises to higher-dimensional observation spaces, i.e.\ $\dim\Y > 1$. Here, $\mathcal{B}_q(x) \subseteq \Y$ specifies a set of predictions which is estimated to contain the mean of observations which are made at the conditions $x \in \X$ with a probability of $q$, which illustrates that it is equivalent to the conventional definition from \cref{eqn:PWConfBands1DDef}. Again, this is to be understood in the frequentist sense that the confidence bands $\mathcal{B}_q(x)$ computed for different dataset realisations envelop the true value $y_\model(x; \theta_\text{true})$ in a fraction $q \in (0,1)$ of realisations for a given $x \in \X$.

%
%

A definition of pointwise confidence bands in this manner also has the benefit of not presupposing any particular form for the uncertainty distribution of the observed data (e.g.\ a normal distribution) around the model. Instead, the effects of any given data uncertainty distribution are already incorporated into the confidence regions $\mathcal{C}_q$ via the likelihood function. Therefore, the confidence bands $\mathcal{B}_q$ remain unaffected by non-linear reparametrisations of models.


Just as with confidence regions, the boundary of a pointwise confidence band $(\partial \mathcal{B}_q)(x) = \partial \pqty\big{y_\model(x;\mathcal{C}_q)}$ is of particular interest for the purpose of illustration. That is, one wishes to draw curves or surfaces which are estimated to encompass the prediction of the true model underlying the data with a confidence level $q$. Incidentally, there exists a class of models for which it suffices to evaluate the model only on the boundary of a confidence region $\partial\mathcal{C}_q$, instead of on the full interior $\mathcal{C}_q$, in order to obtain the boundary of the confidence band $(\partial \mathcal{B}_q)(x)$.



Specifically, for a map $y:\M \longrightarrow C^0(\X,\Y)$ and some set $C \subseteq \M$, one would like to prove the topological relation
\begin{equation} \label{eqn:TopInclusion}
	\partial\pqty\big{y(C)} \subseteq y(\partial C)
\end{equation}
under the weakest assumptions possible. 
A detailed proof is given in \secref[sec:TopologicalProof]{appendix \ref*{sec:TopologicalProof}} which shows that sufficient conditions for \eqnref[eqn:TopInclusion]{relation (\ref*{eqn:TopInclusion})} to hold are that the map be injective as well as continuous and that the set $C \subseteq \M$ be compact. The injectivity and continuity of a model map are given if it is globally structurally identifiable on $C$. An appropriate set $C$ is given by the closure $\overline{\mathcal{C}_q}$ of any bounded confidence region $\mathcal{C}_q$. As discussed in \cref{sec:ParameterIdentifiability}, this boundedness of $\mathcal{C}_q$ is equivalent to the practical identifiability of the model at the confidence level $q$. Moreover, if the Hessian of the likelihood is negative-definite at the MLE, i.e.\ if it constitutes a true maximum, there always exists a $q > 0$ such that the associated confidence region $\mathcal{C}_q$ is bounded.


When applicable, \eqnref[eqn:TopInclusion]{relation (\ref*{eqn:TopInclusion})} represents a considerable reduction in computational effort since sampling of the interior of the confidence region $\mathcal{C}_q$ can be avoided and thus fewer evaluations of the model are necessary to construct the desired confidence bands. In particular, the integral manifold method described in \cref{sec:IsoLikelihoodSurfaces} provides a convenient parametrisation of confidence boundaries $\partial \mathcal{C}_q$ such that the model can be efficiently evaluated on parameters $\theta \in \partial \mathcal{C}_q$ at any desired $x \in \X$ to establish the simultaneous confidence bands. Moreover, for large datasets where the main computational bottleneck is caused by the evaluation of the full likelihood rather than the computation of any individual prediction $y_\model(x;\theta)$, confidence bands $\mathcal{B}_q$ are obtained with little additional computational effort once the associated confidence regions $\mathcal{C}_q$ are known.



It should be stressed that confidence bands around the best fit prediction are not a reflection of how well the model predictions agree with the observed data. Instead, they demonstrate how the uncertainties in the parameters propagate to the predictions of a model and thereby illustrate the flexibility inherent in the model. Furthermore, under the assumption that the given model indeed provides the correct description of the observational data, the confidence bands constitute a faithful assessment of the probability of covering the model prediction associated with the true parameter configuration.

Lastly, since the confidence bands are wider for $x$-values where the uncertainty in the model predictions is larger, one can use their size $\mathrm{vol}\pqty{\mathcal{B}_q(x)\vphantom{^3}}$ to judge under which conditions new observations will contribute the highest amount of useful information to constrain the model predictions. Therefore, confidence bands also serve as a useful tool in the design of experiments.

\subsection{Effects of Non-Linear Reparametrisations on Confidence Regions} \label{sec:Survey} \FigureNames{Survey}

\begin{figure}[!ht]
	\centering
	\begin{subfigure}{\textwidth}
	\begin{minipage}[c]{.77\textwidth}
		\centering
		\PublishGraphic[Saved]{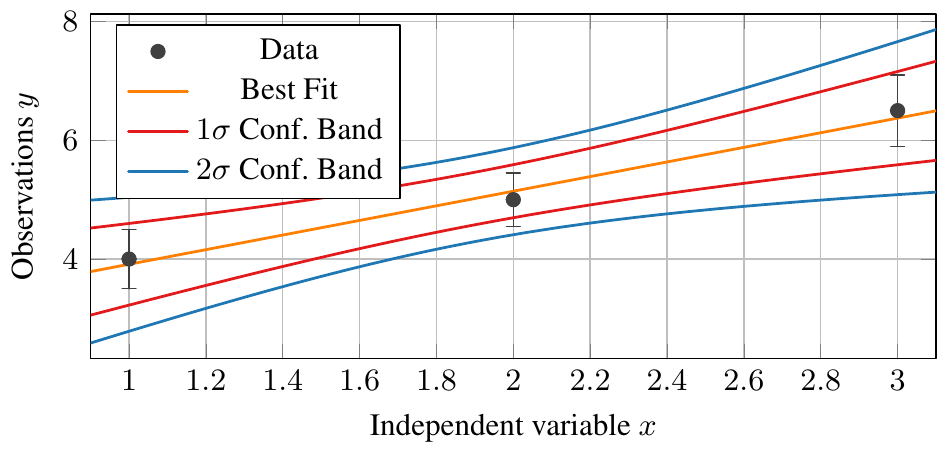}{\begin{tikzpicture}
				\pgfplotstableread[col sep=comma]{PlotData/ModelTiles/ConfBands1.txt}{\CBa}
				\pgfplotstableread[col sep=comma]{PlotData/ModelTiles/ConfBands2.txt}{\CBb}
				\begin{axis}[xlabel={Independent variable $x$},ylabel={Observations $y$}, legend style={font=\footnotesize}, clip mode=individual, legend pos=north west, small, width=0.8\textwidth, height=0.4\textwidth]
					\addplot [plotgrey, only marks, error bars/.cd, x dir=both, x explicit, y dir=both, y explicit] coordinates {
						(1,4) +- (0,0.5)
						(2,5) +- (0,0.45)
						(3,6.5) +- (0,0.6)};
					\addplot+[fitcol,samples=2,domain=0.9:3.1] {1.23063*x+2.68134};
					\addplot+[forget plot,mycol1,smooth] table [x index = 0, y index = 1] {\CBa};   \addplot+[mycol1,smooth] table [x index = 0, y index = 2] {\CBa};
					\addplot+[forget plot,mycol2,smooth] table [x index = 0, y index = 1] {\CBb};   \addplot+[mycol2,smooth] table [x index = 0, y index = 2] {\CBb};
					\legend{Data,Best Fit,$1\sigma$ Conf.~Band, $2\sigma$ Conf.~Band}
				\end{axis}
		\end{tikzpicture}}
	\end{minipage}
	\begin{minipage}[c]{.22\textwidth}
		\centering
		\begin{tabular}{c c c}
			\toprule
			$x$ & $y$ & $\sigma$\\
			\midrule
			$1$ & $4$ & $0.5$\\
			$2$ & $5$ & $0.45$\\
			$3$ & $6.5$ & $0.6$\\
			\bottomrule
		\end{tabular}
	\end{minipage}
	\caption{A brief summary of the dataset on which the confidence regions from \cref{fig:Reparametrisations} are based. In addition to the data and best fit, the left-hand side plot depicts the pointwise confidence bands of level $1\sigma$ and $2\sigma$ generated from the confidence boundaries.}
	\label{fig:ReparametrisationData}
	\end{subfigure}\\\vspace{0.65em}
	\begin{subfigure}{\textwidth}
	\PublishGraphic[Saved]{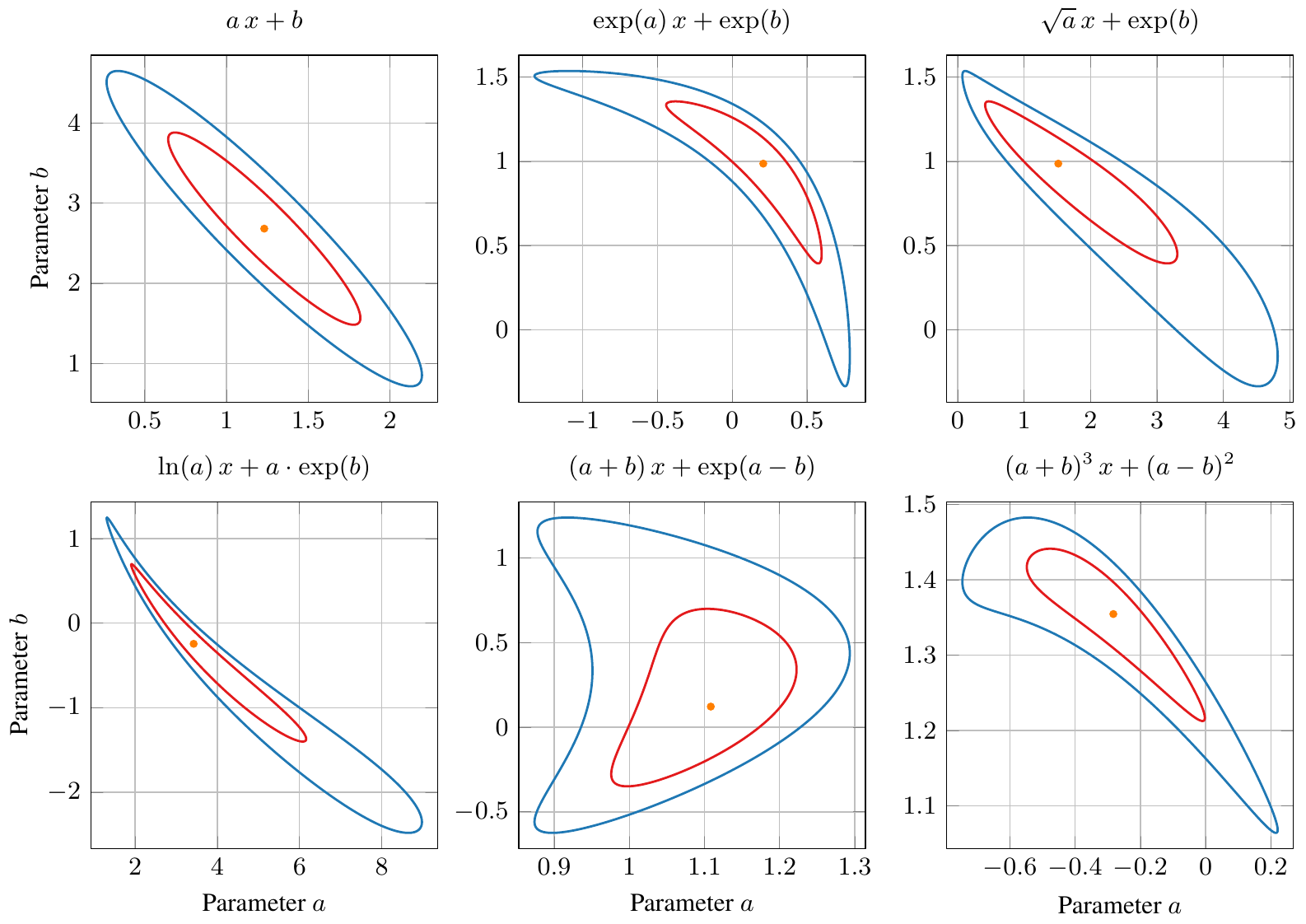}{\begin{tikzpicture}
			\pgfplotstableread[col sep=comma]{PlotData/ModelTiles/model1.txt}{\MAa}
			\pgfplotstableread[col sep=comma]{PlotData/ModelTiles/model6.txt}{\MAf}
			\pgfplotstableread[col sep=comma]{PlotData/ModelTiles/model7.txt}{\MAg}
			\pgfplotstableread[col sep=comma]{PlotData/ModelTiles/model8.txt}{\MAh}
			\pgfplotstableread[col sep=comma]{PlotData/ModelTiles/model10.txt}{\MAj}
			\pgfplotstableread[col sep=comma]{PlotData/ModelTiles/model12.txt}{\MAl}
			\begin{groupplot}[group style={group size=3 by 2, vertical sep=3.5em},enlarge x limits = 0.05, small,width=0.355\textwidth,height=0.355\textwidth,clip=false]
				\nextgroupplot[title={$\footnotesize\smash{a \, x + b}$\vphantom{I}}, ylabel={Parameter $b$}]
				\addplot+[smooth] table [x index=0, y index=1] {\MAa} --cycle;
				\addplot+[smooth] table [x index=2, y index=3] {\MAa} --cycle;
				\node [mpoint] at (axis cs:1.23063, 2.68134) {};
				\nextgroupplot[title={$\footnotesize\smash{\exp(a) \, x + \exp(b)}$}]
				\addplot+[smooth] table [x index=0, y index=1] {\MAf} --cycle;
				\addplot+[smooth] table [x index=2, y index=3] {\MAf} --cycle;
				\node [mpoint] at (axis cs:0.207529, 0.986316) {};
				\nextgroupplot[title={$\footnotesize\smash{\sqrt{a} \, x + \exp(b)}$}]
				\addplot+[smooth] table [x index=0, y index=1] {\MAg} --cycle;
				\addplot+[smooth] table [x index=2, y index=3] {\MAg} --cycle;
				\node [mpoint] at (axis cs:1.51446, 0.986316) {};
				\nextgroupplot[title={$\footnotesize\smash{\ln(a) \, x + a\cdot\exp(b)}$}, ylabel={Parameter $b$}, xlabel={Parameter $a$}]
				\addplot+[smooth] table [x index=0, y index=1] {\MAh} --cycle;
				\addplot+[smooth] table [x index=2, y index=3] {\MAh} --cycle;
				\node [mpoint] at (axis cs:3.4234, -0.244318) {};
				\nextgroupplot[title={$\footnotesize\smash{(a+b) \, x + \exp(a-b)}$}, xlabel={Parameter $a$}]
				\addplot+[smooth] table [x index=0, y index=1] {\MAj} --cycle;
				\addplot+[smooth] table [x index=2, y index=3] {\MAj} --cycle;
				\node [mpoint] at (axis cs:1.10847, 0.122159) {};
				\nextgroupplot[title={$\footnotesize\smash{(a+b)^3 \, x + (a-b)^2}$}, xlabel={Parameter $a$}]
				\addplot+[smooth] table [x index=0, y index=1] {\MAl} --cycle;
				\addplot+[smooth] table [x index=2, y index=3] {\MAl} --cycle;
				\node [mpoint] at (axis cs:-0.282927, 1.35455) {};
			\end{groupplot}
	\end{tikzpicture}}
	\caption{Illustration of confidence regions of levels $1\sigma$ and $2\sigma$ for various alternative parametrisations of a model function with parameters $\theta = (a,b) \in \R^2$. For each of the chosen parametrisations, the model prediction consists of a straight line which is fitted to the dataset shown in \cref{fig:ReparametrisationData}.}
	\label{fig:Reparametrisations}
	\end{subfigure}
	\caption{Using a toy dataset, this survey on the effects of non-linear model reparametrisation on confidence regions demonstrates that even relatively simple algebraic manipulations of the model can induce strong distortions in the resulting confidence regions.}
\end{figure}
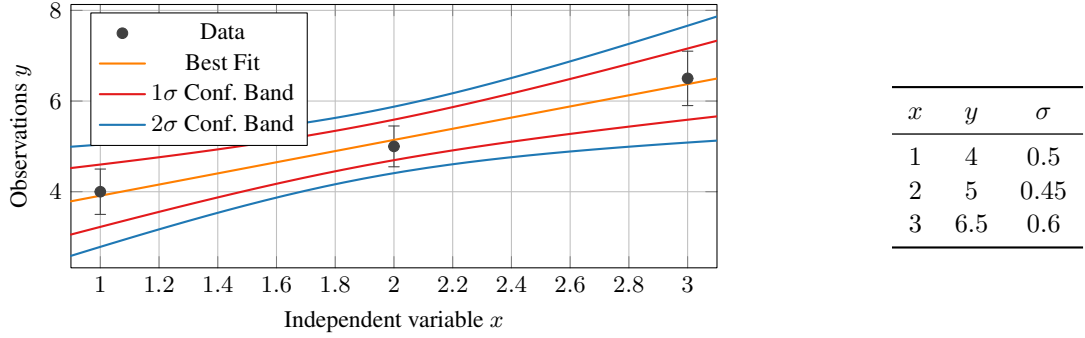
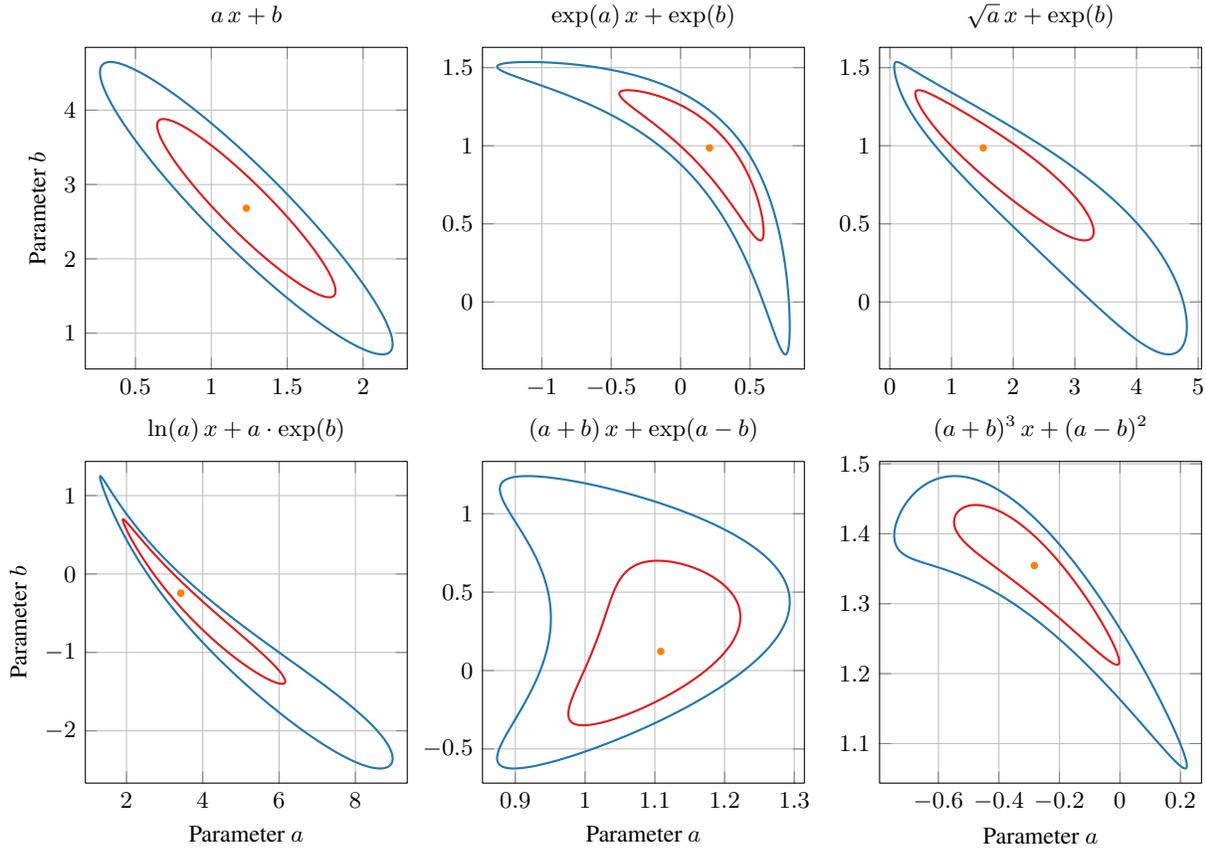

The aim of this section is to provide a small survey which illustrates the qualitative effects that model reparametrisations can have on the shapes of confidence regions. Discussions of the suitability of the employed parametrisations such as their invertibility, differentiability, valid chart domains and so on are omitted in these examples and assumed not to pose any technical issues. The analysed toy dataset consists of only three observations and is illustrated in \cref{fig:ReparametrisationData}. Each of the various model parametrisations shown in \cref{fig:Reparametrisations} correspond to the choice of a different chart on the same embedded prediction surface in the data space $h(\M) \subseteq \Y^N$, which encodes a linear relationship between $\X$ and $\Y$ in this case. As a result, while the confidence regions $\mathcal{C}_q$ for these parametrisations exhibit different coordinate distortions, their image under the corresponding embedding map $h(\mathcal{C}_q)$ is the same.

\Cref{fig:Reparametrisations} demonstrates the impact of non-linearity in model parametrisations on the location, size and shape of confidence boundaries and the apparent coordinate distortion of the parameter space in general. It also reveals that the deviations in the shapes of confidence regions from perfect ellipsoids generally increase with confidence level, i.e.\ with radial coordinate distance from the MLE. That is, the approximation of the confidence boundaries as ellipsoids generally becomes worse with increasing confidence level. Since confidence regions of differing levels are no longer similar in the mathematical sense that there exists a uniform scaling factor which makes them congruent, they have to be computed individually for each confidence level of interest. This highlights a further conceptual weakness of using a covariance matrix to approximate non-linear parameter uncertainties, namely that there is no clear-cut way of assessing the magnitude of the non-linear distortion of any given confidence region a-priori and consequently how well the exact confidence region is approximated by an ellipsoid. 

Although this toy model and its various non-linear reparametrisations which are explored here constitute a somewhat artificial example, it is important to keep in mind that throughout many scientific disciplines, models borne out of theoretical considerations are generically non-linear with respect to their parameters. That is, non-linearity with respect to the model parameters should arguably be regarded as the typical case.

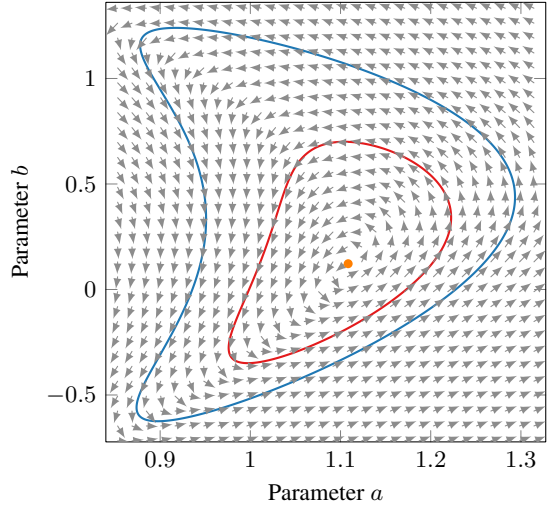
\begin{figure}[!ht]
	\centering
	\PublishGraphic[Saved]{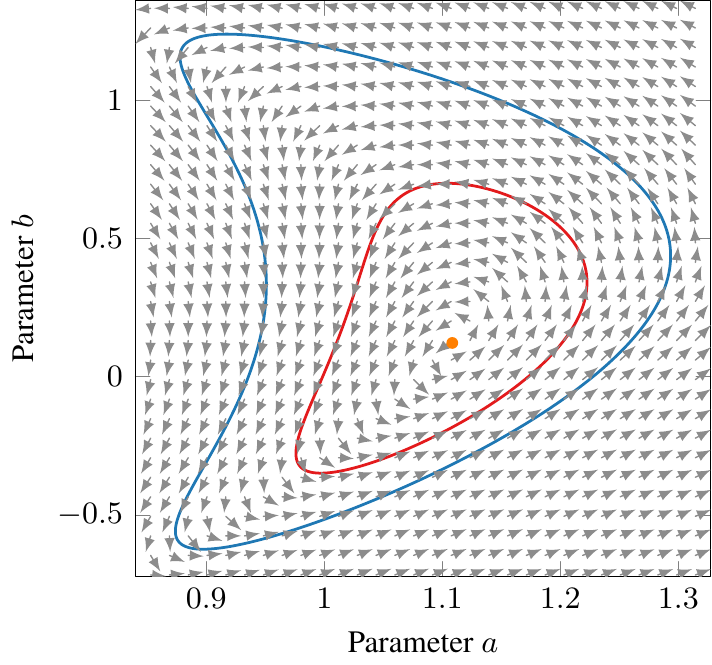}{\begin{tikzpicture}
		\pgfplotstableread[col sep=comma]{PlotData/NonLinearToyOrthVF.txt}{\Arrows}
		\pgfplotstableread[col sep=comma]{PlotData/ModelTiles/model10.txt}{\Confn}
		\begin{axis}[enlargelimits=false, clip mode = individual, grid=none, xlabel={Parameter $a$}, ylabel={Parameter $b$}, small, width=.45\textwidth, height=.45\textwidth, ytick distance=0.5]
			\addplot+[z filter/.expression={10}] table [x index=0, y index=1] {\Confn} --cycle;
			\addplot+[z filter/.expression={10}] table [x index=2, y index=3] {\Confn} --cycle;
			\addplot [gray!90!white, quiver={u=\thisrow{u},v=\thisrow{v}},-{latex[length=2pt]}] table {\Arrows};
			\node [mpoint, inner sep=1.2pt] at (axis cs:1.10847, 0.122159) {};
		\end{axis}
	\end{tikzpicture}}
	\caption{Illustration of a suitably normalised likelihood-annihilating vector field generated for the model parametrisation $y_\model\pqty{\vphantom{(^1}x;(a,b)} \coloneqq (a+b)\,x + \exp(a-b)$ and the dataset shown in \cref{fig:ReparametrisationData} together with the $1\sigma$ and $2\sigma$ confidence boundaries.}
	\label{fig:OrthVFPlot}
\end{figure}

\Cref{fig:OrthVFPlot} shows an example of likelihood-annihilating vector fields which have been constructed according to \cref{eqn:OrthVF}, along with the integral curves they generate when given appropriate initial conditions. The fact that integrating along the illustrated vector field indeed results in closed curves is indicative of the stability and accuracy of this scheme. For further details, see \cref{sec:Performance}.

\subsection{Applications}

In the following, we aim to demonstrate how knowledge of the exact confidence regions can afford insights into real-world problems. The first example is taken from fundamental physics and investigates the relationship between the apparent distance and redshift of objects under the assumption of a flat cosmological spacetime filled with matter and dark energy. The second example is taken from the subject of systems biology, where models are often defined implicitly via the solution to a system of differential equations.

%
%


\FigureNames{Cosmology}

\subsubsection{Distance--Redshift Relationship of Type Ia Supernov\ae} \label{sec:SCPAnalysis}


The Supernova Cosmology Project (SCP) dataset which is used in the following contains \num{580} independent measurements of distant type Ia supernov\ae\ and is publicly available \cite{SCPData}. An analysis of this dataset by conventional methods can be found e.g.\ in \cite{SCPPaper}.

One possible way of quantifying distance in a cosmological setting is via the so-called distance modulus $\mu$ which can be expressed as a function of the cosmological redshift $z$ by
\begin{equation} \label{eqn:DistModulus}
	\mu(z;\Omega_\text{m},w) = 10 + 5 \log_{10}\!\pqty{(1+z) \, d_\text{H} \int_{0}^{z} \! \dd{x} \, \frac{1}{\sqrt{\Omega_\text{m} \, (1+x)^3 + (1-\Omega_\text{m}) (1+x)^{3(1+w)}}}}
\end{equation}
where $d_\text{H} = c/H_0$ is the Hubble distance, $\Omega_\text{m}$ is the matter density in the Universe as observed today and $w$ is the dark energy equation of state parameter today. As indicated by the notation $\mu(z;\Omega_\text{m},w)$, the model is interpreted as having two variable parameters $\Omega_\text{m}$ and $w$, whereas the Hubble--Lema{\^i}tre constant is fixed at an assumed value of $H_0 = \SI{70}{\frac{km}{s \cdot Mpc}}$ for the sake of this example.


In this particular case, numerical integration can be avoided since there exists a closed form solution for the integral in \cref{eqn:DistModulus} which can be found by symbolic integration algorithms to be
\begin{equation}
	\int \! \dd{u} \frac{1}{\sqrt{A \, u^3 + B \, u^c}} = -\frac{2 u \sqrt{\frac{A \, u^{3 - c}}{B} + 1} \, \cdot {}_{2}F_{1}\pqty{\frac{1}{2}, \frac{c - 2}{2c - 6}; \frac{3 c - 8}{2 c - 6}; -\frac{A \, u^{3 - c}}{B}}}{(c - 2) \sqrt{A \, u^3 + B \, u^c}} + \text{const.}
\end{equation}
where ${}_2F_1$ is the hypergeometric function defined by
\begingroup\allowdisplaybreaks
\begin{align}
	{}_{2}F_{1}(a,b;c;z) 
	= \frac{\Gamma (c)}{\Gamma (b) \,\Gamma (c-b)} \, \int_{0}^{1} \! \dd{t} t^{b-1} \, (1-t)^{c-b-1} \, (1-zt)^{-a}.
\end{align}\endgroup%
Thus, existing approximations of the hypergeometric function ${}_{2}F_{1}$ can be used to efficiently compute solutions to the definite integral in \cref{eqn:DistModulus} which not only reduces the overall computational effort significantly, but also increases the accuracy compared with direct numerical integration schemes. By exploiting the differentiability of the distance modulus $\mu(z;\Omega_\text{m},w)$ with respect to the redshift $z$, one can verify that this model is injective for $w \neq 0$ and $\Omega_\text{m} \neq 1$, i.e.\ on the domain $\M = \,\Setcond{(\Omega_\text{m},w) \in \R^2}{0 < \Omega_\text{m} < 1, w < 0}$. 

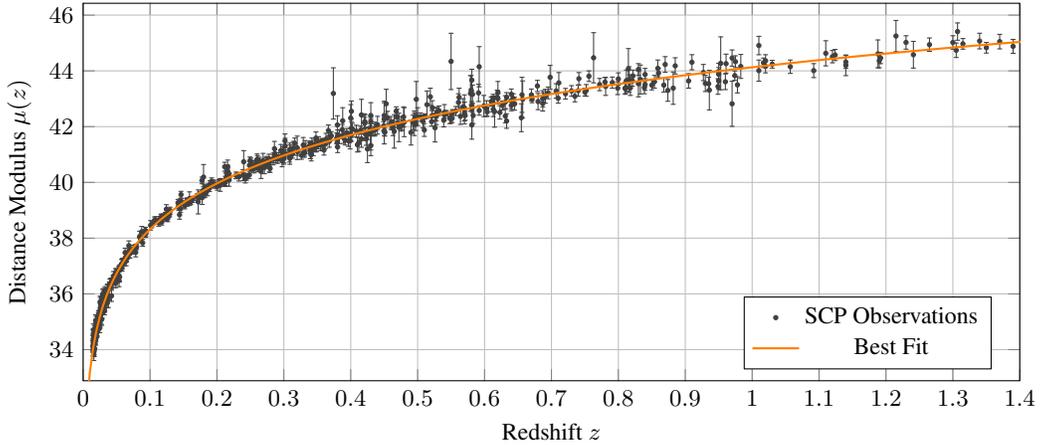
\begin{figure}[!ht]
	\centering
	\PublishGraphic[Saved]{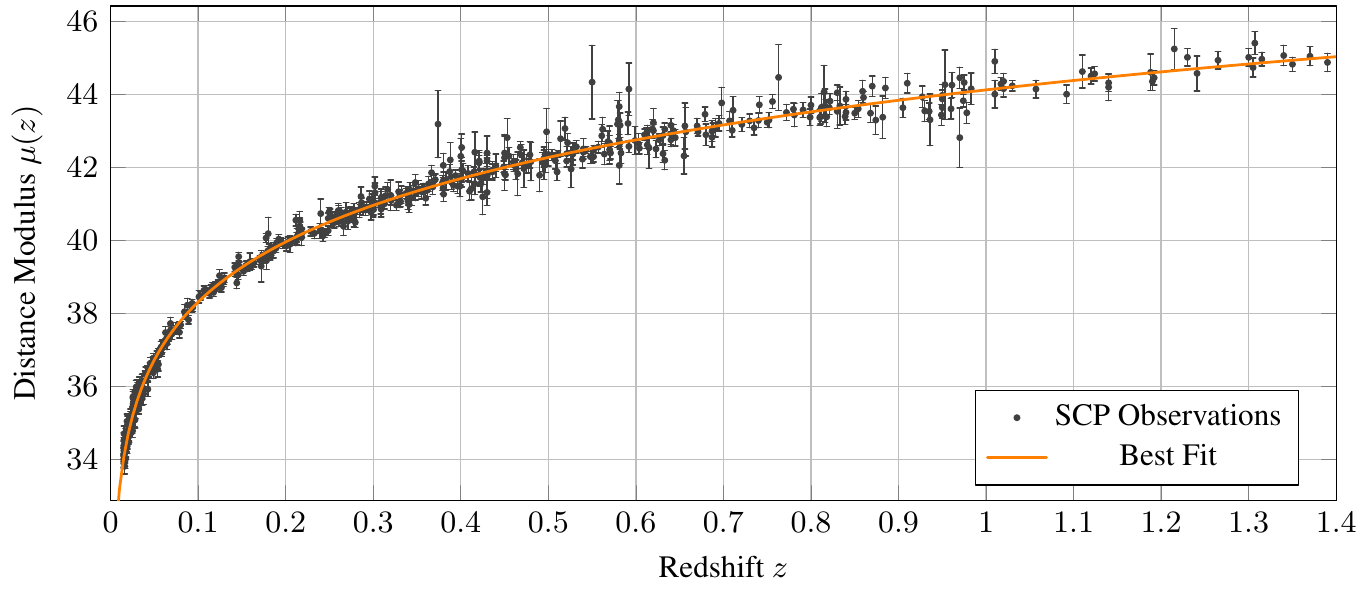}{\begin{tikzpicture}
		\begin{axis}[xlabel={Redshift $z$}, ylabel={Distance Modulus $\mu(z)$}, legend pos=south east, xmin=0, xmax=1.4, ymin = 33.5, clip mode=individual, legend style={font=\footnotesize}, small, width=0.85\textwidth, height=0.4\textwidth]
			\pgfplotstableread[col sep=comma]{PlotData/SCPData.txt}{\SCP}
			\pgfplotstableread[col sep=comma]{PlotData/SCPDataBestFit.txt}{\SCPFit}
			\addplot[plotgrey, mark size=0.8pt, only marks, error bars/.cd, x dir=both, x explicit, y dir=both, y explicit] table [x index=0, y index=1, y error index=2] {\SCP};
			\addplot+[thick, fitcol] table {\SCPFit};
			\legend{SCP Observations,Best Fit}
		\end{axis}
	\end{tikzpicture}}
	\caption{Visualisation of the distance modulus $\mu(z)$ as a function of redshift $z$ for observations of type Ia supernov\ae, measured by the Supernova Cosmology Project. A fit of \cref{eqn:DistModulus} to the data was performed using maximum likelihood estimation to determine the optimal parameters of $\Omega_\text{m} \approx \num{0.28}$ and $w \approx \num{-1.00}$ under the assumption of $H_0 = \SI{70}{\frac{km}{s \cdot Mpc}}$. The data was excerpted from \cite{SCPData}.}
	\label{fig:SCPData}
\end{figure}

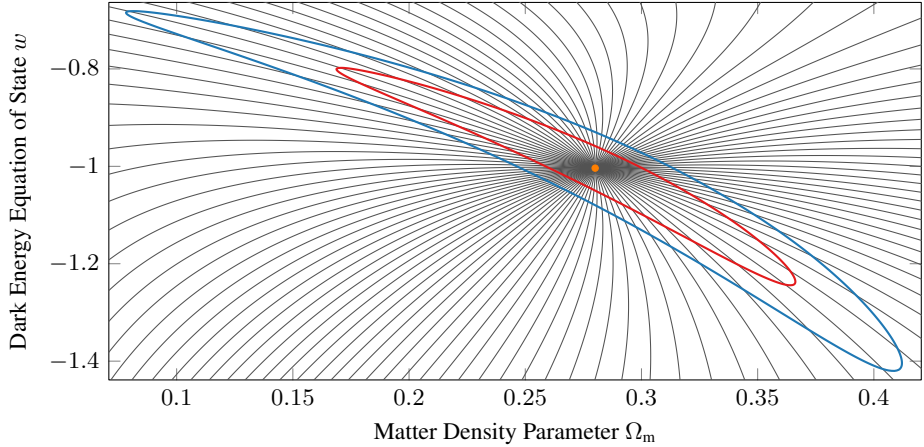
\begin{figure}[!ht]
	\centering
	\PublishGraphic[Saved]{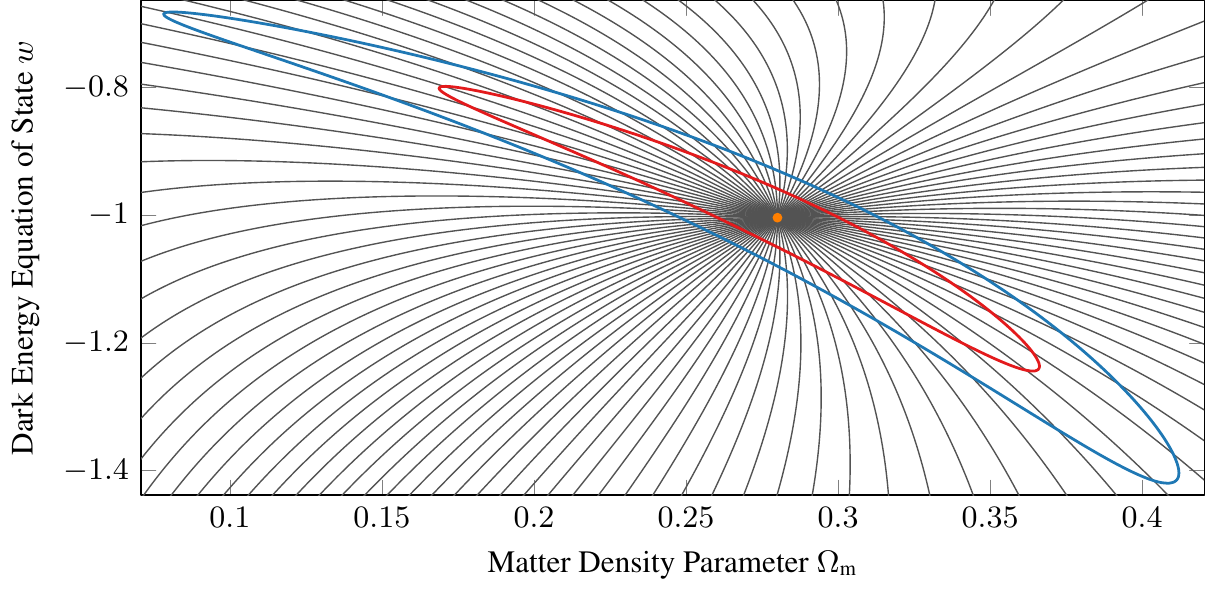}{\begin{tikzpicture}
		\pgfplotstableread[col sep=comma]{PlotData/SCPConf1sigma.txt}{\ConfONE}
		\pgfplotstableread[col sep=comma]{PlotData/SCPConf2sigma.txt}{\ConfTWO}
		\pgfplotstableread[col sep=comma]{PlotData/SCPgeos.txt}{\Geos}
		\begin{axis}[xlabel={Matter Density Parameter $\Omega_\text{m}$},ylabel={Dark Energy Equation of State $w$}, 
		enlargelimits=false, grid=none, xmin=0.0708140, ymin=-1.4385646, xmax=0.4204016, ymax=-0.6634777, xtick distance=0.05, legend style={font=\footnotesize}, small, width=0.75\textwidth, height=0.4\textwidth]
			\foreach \n in {0,2,...,198}{
				\pgfmathsetmacro{\m}{\n+1}
				\addplot [thin,plotgrey!90!white] table [x index = \n, y index=\m] {\Geos};}
			\addplot+[smooth,draw=mycol2
			] table [x index = 0, y index=1] {\ConfTWO};
			\addplot+[smooth,draw=mycol1
			] table [x index = 0, y index=1] {\ConfONE};
			\addplot [fitcol,only marks, mark size=1.2pt] coordinates {(0.28007,-1.00369)};
		\end{axis}
	\end{tikzpicture}}
	\caption{Plot of the exact $1\sigma$ and $2\sigma$ confidence regions together with $100$ radial geodesics emanating from the MLE. Since geodesics constitute straight lines on the underlying parameter manifold, their curved appearance in the given coordinate chart illustrates the magnitude of the effective parameter non-linearity in the distance modulus model.
	}
	\label{fig:SCPConfidenceRegion}
\end{figure}

The SCP dataset is illustrated in \cref{fig:SCPData} together with the distance modulus from \cref{eqn:DistModulus} evaluated at the maximum likelihood estimate $(\Omega_\text{m},w) \approx (0.28,-1.00)$. 
Especially from \cref{fig:SCPConfidenceRegion}, it is evident that the iso-likelihood contours are of non-ellipsoidal shape which illustrates the non-linearity of the distance modulus $\mu$ with respect to $\Omega_\text{m}$ and $w$, which can also be seen from \cref{eqn:DistModulus}. In addition, the radial geodesics depicted in \cref{fig:SCPConfidenceRegion} provide a visual indication of the non-linear coordinate distortion which is present on the manifold via their curved shapes.

By conditionalising $\Omega_\text{m}$ to a very small value and maximising the likelihood with respect to the remaining parameter $w$, one finds that the largest confidence region which does not yet intersect the $\Omega_\text{m} = 0$ boundary, below which the distance modulus model in its parametrisation from \cref{eqn:DistModulus} no longer provides a valid description, is approximately of level $q \approx \num{2.73}\sigma \approx \SI{99.4}{\%}$. The bent shapes of the exact confidence boundaries depicted in \cref{fig:SCPConfidenceRegion} show that at low values of $\Omega_\text{m}$, the range of likely values for $w$ not only becomes more constrained but that the model also appears to get less sensitive towards changes in $\Omega_\text{m}$.


\begin{figure}[!ht]
	\centering
	\PublishGraphic[Saved]{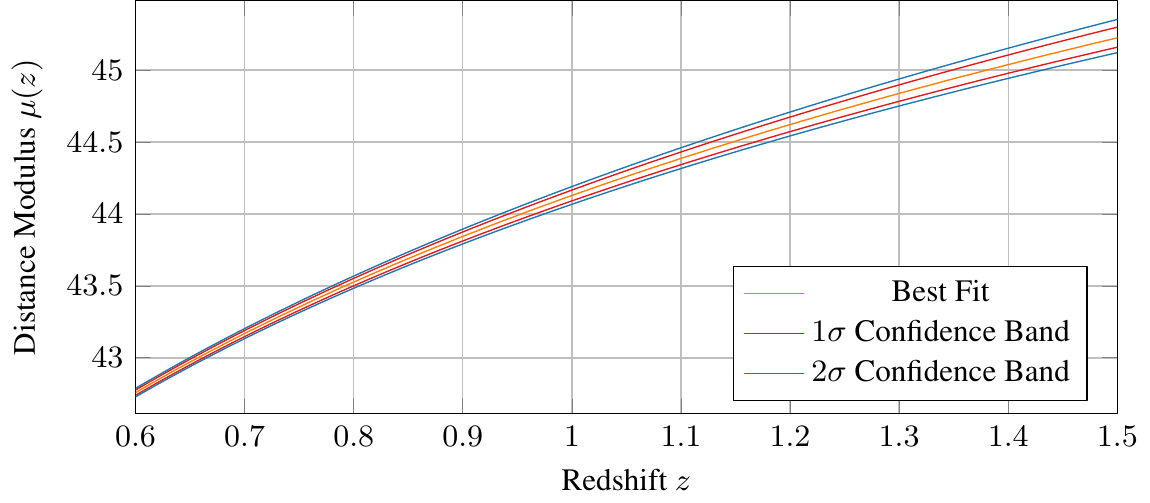}{\begin{tikzpicture}
		\pgfplotstableread[col sep=comma]{PlotData/SCPDataBestFit.txt}{\SCPFit}
		\pgfplotstableread[col sep=comma]{PlotData/SCPConfBand1.txt}{\SCPBandOne}
		\pgfplotstableread[col sep=comma]{PlotData/SCPConfBand2.txt}{\SCPBandTwo}
		\begin{axis}[xlabel={Redshift $z$}, ylabel={Distance Modulus $\mu(z)$},enlarge x limits=false, legend pos = south east, xmin=0.6, xmax=1.5, style={font=\footnotesize}, small, width=0.7\textwidth, height=0.35\textwidth]
			\begin{scope}[thin]
				\addplot+[fitcol] table {\SCPFit};
				\addplot+[mycol1] table [x index = 0, y index=1] {\SCPBandOne};
				\addplot+[mycol1,forget plot] table [x index = 0, y index=2] {\SCPBandOne};
				\addplot+[mycol2] table [x index = 0, y index=1] {\SCPBandTwo};
				\addplot+[mycol2,forget plot] table [x index = 0, y index=2] {\SCPBandTwo};
			\end{scope}
			\legend{Best Fit,$1\sigma$ Confidence Band,$2\sigma$ Confidence Band}
		\end{axis}
	\end{tikzpicture}}
	\caption{Plot of pointwise confidence bands for the SCP dataset and distance modulus model around the maximum likelihood prediction using the method described in \cref{sec:ConfidenceBands}. Evidently, the model prediction is strongly constrained by the available data for small and medium redshifts $z$ whereas uncertainty in the model prediction increases for higher redshift. This is an indication that further observations at high redshift would best constrain the model prediction.}
	\label{fig:SCPConfidenceBands}
\end{figure}
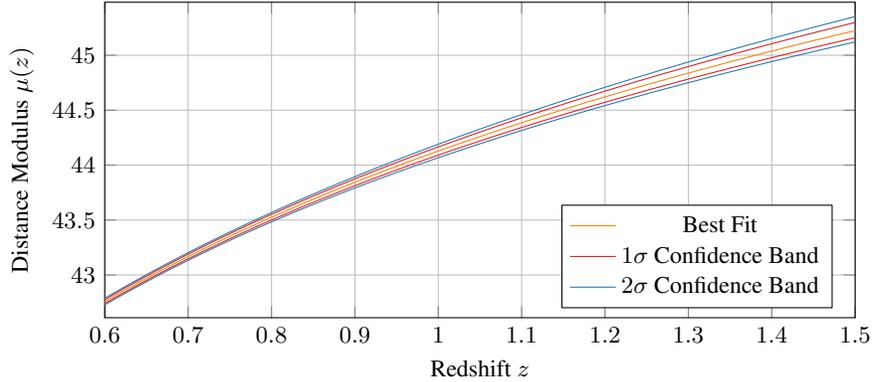

Lastly, \cref{fig:SCPConfidenceBands} depicts the $1\sigma$ and $2\sigma$ confidence bands associated with the maximum likelihood prediction of the distance modulus model, whose widths increase for higher redshifts $z$. This is an indication that observations at high redshifts contain the most amount of useful information about the model parameters, which is consistent with what one would expect from the underlying physical theory.

\subsubsection{Modelling of Infectious Diseases} \label{sec:DiseaseModelling}

\FigureNames{Biology}


One of the simplest models for describing the spread of an infectious disease is the so-called SIR model \cite{ToensingInfection}. In essence, it divides the total population into susceptible, infected and recovered sub-populations where the rates of infection and recovery are controlled by two parameters $\beta, \gamma \in \R^+$. The model is characterised by 
the system of ODEs given by
\begin{equation} \label{eqn:SIRode}
\dv{S(t)}{t} = - \beta \, S(t) \, I(t) \qqc
\dv{I(t)}{t} = \beta \, S(t) \, I(t) - \gamma \, I(t) \qqc
\dv{R(t)}{t} = \gamma \, I(t).
\end{equation}
ODE-based approaches such as this assume that the sub-populations are large enough to be modelled as real numbers and well-mixed. While this basic SIR model is certainly an oversimplification of the mechanisms underlying any real-world outbreaks of infectious diseases, there are various ways to extend this model such that it provides a more accurate description of real disease transmission, for instance by allowing for time dependence in the parameters \cite{SIRpaper, SpoCK}. 

\begin{table}[!ht]
	\centering
	\begin{small}
		{\def\arraystretch{1.6}%
			\begin{tabular}{c c c c c c c c c c c c c c c}
				\toprule
				Time $t$ \un{days} & 1 & 2 & 3 & 4 & 5 & 6 & 7 & 8 & 9 & 10 & 11 & 12 & 13 & 14\\
				\# Infected & 3 & 8 & 28 & 75 & 221 & 291 & 255 & 235 & 190 & 126 & 70 & 28 & 12 & 5\\
				\bottomrule
		\end{tabular}}
		\vspace{0.6em}
	\end{small}
	\caption{Dataset from an influenza outbreak at an English boarding school in 1978, reproduced from \cite{SIRpaper}. Since the original publication does not cite any uncertainties in the number of infected pupils, we will assume the uncertainties to be $\sigma = 15$ in all calculations for sake of simplicity. The total number of pupils at the school is reported as \num{763}.}
	\label{tab:BoardingSchoolData}
\end{table}

\begin{figure}[!ht]
	\centering
	\PublishGraphic[Saved]{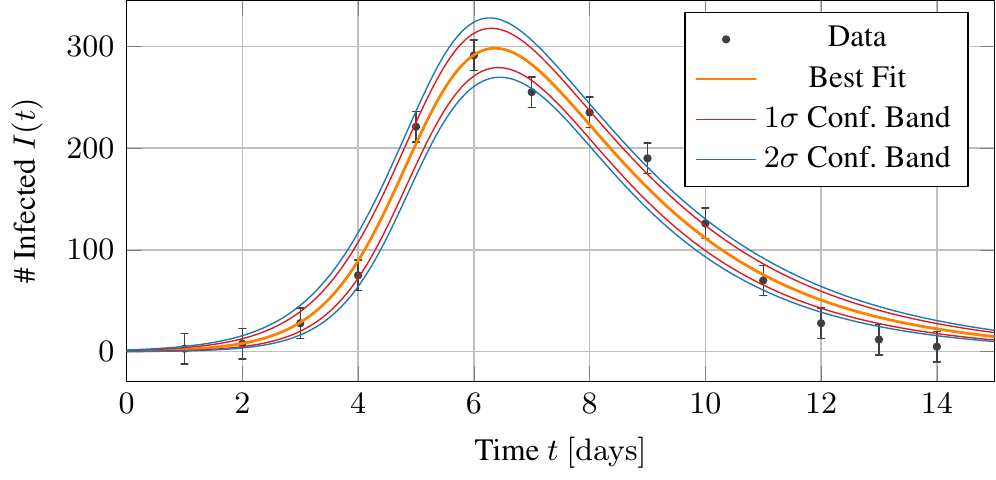}{\begin{tikzpicture}
		\begin{axis}[xlabel={Time $t$ \un{days}}, ylabel={\# Infected $I(t)$}, legend pos=north east, xmin=0, xmax=15, clip mode=individual, style={font=\footnotesize}, small, width=0.63\textwidth, height=0.33\textwidth]
			\pgfplotstableread[col sep=comma]{PlotData/SIRdata.txt}{\SIR}
			\pgfplotstableread[col sep=comma]{PlotData/SIR3Dfit.txt}{\SIRfit}
			\pgfplotstableread[col sep=comma]{PlotData/SIR3Dbands1.txt}{\SIRbandsI}
			\pgfplotstableread[col sep=comma]{PlotData/SIR3Dbands2.txt}{\SIRbandsII}
			\addplot[plotgrey, mark size=1pt, only marks, error bars/.cd, x dir=both, x explicit, y dir=both, y explicit] table [x index=0, y index=1, y error index=2] {\SIR};
			\addplot+[fitcol, thick] table {\SIRfit};
			\addplot+[thin, mycol1, smooth] table [x index=0, y index=1] {\SIRbandsI};
			\addplot+[thin, mycol1, smooth, forget plot] table [x index=0, y index=2] {\SIRbandsI};
			\addplot+[thin, mycol2, smooth] table [x index=0, y index=1] {\SIRbandsII};
			\addplot+[thin, mycol2, smooth, forget plot] table [x index=0, y index=2] {\SIRbandsII};
			\legend{Data, Best Fit, $1\sigma$ Conf.~Band, $2\sigma$ Conf.~Band}
		\end{axis}
	\end{tikzpicture}}
	\caption{Visualisation of the English Boarding School infection dataset from \cref{tab:BoardingSchoolData} together with best the fit for the parameter triple $\theta = (I_0, \beta, \gamma)$ produced by $\theta_\text{MLE} \approx (0.614, 0.00231, 0.458)$ and the associated confidence bands of levels $1\sigma$ and $2\sigma$, which illustrate the flexibility in the model.}
	\label{fig:SIRdata}
\end{figure}
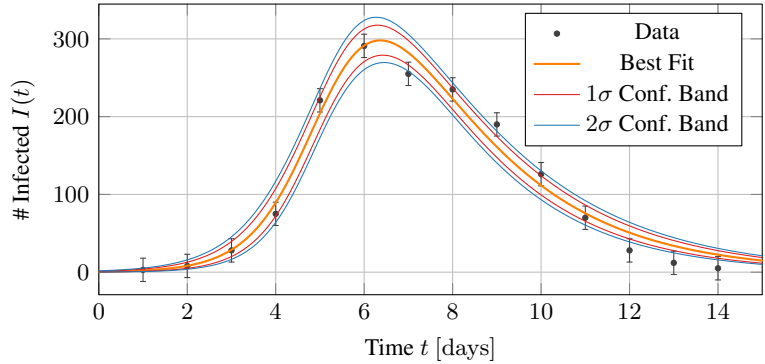

A summary of the dataset used in this example is given in \cref{tab:BoardingSchoolData}. As is often the case, the initial conditions for the ODE system underlying the model are not precisely known. In this particular case, it is unknown at what point in time the initial infection took place or even whether the disease was introduced to the school by multiple students simultaneously. Therefore, the initial number of infected pupils $I_0 \in \R^+$ on day zero is included as an additional parameter in the model and estimated from data. Given that the total number of pupils is reportedly $763$, this yields the constraint $763 \overset{!}{=} S(t) + I(t) + R(t)$ and assuming further that $R(0) = 0$, the initial conditions finally work out to
\begin{equation}
	\pqty\big{S(0), I(0), R(0)} = (763-I_0, I_0, 0).
\end{equation}
Given the mechanistically simple nature of the SIR model, it can be rigorously proven to be injective (i.e.\ globally structurally identifiable) with respect to $\beta$ and $\gamma$, for instance using techniques from differential algebra \cite{SIRidentifiability}. Moreover, in view of the irreversibility of the involved reactions, the injectivity of the SIR model remains unaffected by an inclusion of $I_0$ as a further parameter. 
A visualisation of the 3-parameter SIR model as applied to the dataset from \cref{tab:BoardingSchoolData} is shown in \cref{fig:SIRdata}.





From the confidence bands in \cref{fig:SIRdata}, it is evident that a quantification of the prediction uncertainty via pointwise confidence bands goes beyond a simplistic assumption of the uncertainty in the predictions as being symmetric around the best fit. Instead, it provides a faithful assessment of the varying flexibility of the model across different values of the independent variable $t$. 
One can deduct from the widths of the bands that observations near the peak of the infection wave would be best-suited to further constrain the model parameters. Conversely, measurements which are taken at the very beginning or towards the end of an infection wave, where the susceptible population $S(t)$ is small compared to the total population, contain little useful information about the infection process.

\begin{figure}[!ht]
	\centering
	\PublishGraphic[Saved]{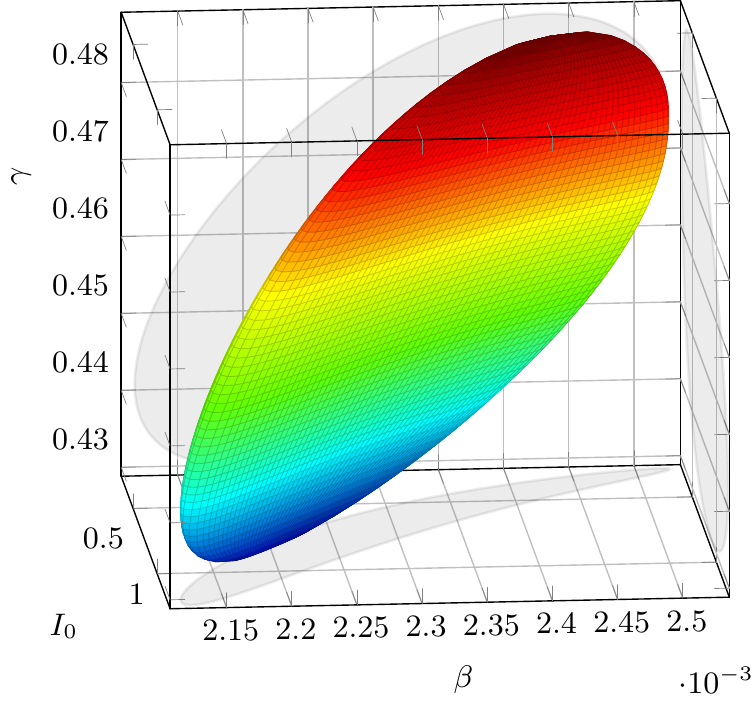}{%
		\begin{tikzpicture}
			\begin{axis}[view={85}{16}, xlabel={$I_0$}, ylabel={$\beta$}, zlabel={$\gamma$}, enlargelimits=false, legend style={font=\footnotesize}, small, width=.47\textwidth, height=.47\textwidth, colormap/bluered,
				3d box=complete, xmin=0.2533648, xmax=1.268026, ymin=0.002106662, ymax=0.002535686, zmin=0.428833, zmax=0.4891406,
				]
				\addplot3[thick, draw=black, fill=black, opacity=0.075, fill opacity=0.075] table [x index=0, y index=1, z expr ={0.428833}, col sep=comma] {PlotData/SIRshadow12.txt} --cycle;
				\addplot3[thick, draw=black, fill=black, opacity=0.075, fill opacity=0.075] table [x index=0, y expr={0.002535686}, z index=1, col sep=comma] {PlotData/SIRshadow13.txt} --cycle;
				\addplot3[thick, draw=black, fill=black, opacity=0.075, fill opacity=0.075] table [x expr={0.2533648}, y index=0, z index=1, col sep=comma] {PlotData/SIRshadow23.txt} --cycle;
				\addplot3[patch, patch type=bilinear, shader=faceted, patch table={PlotData/Faces60.txt}, very thin] file {PlotData/Points60.txt};
			\end{axis}
	\end{tikzpicture}}
	\hfill %
	\PublishGraphic[Saved]{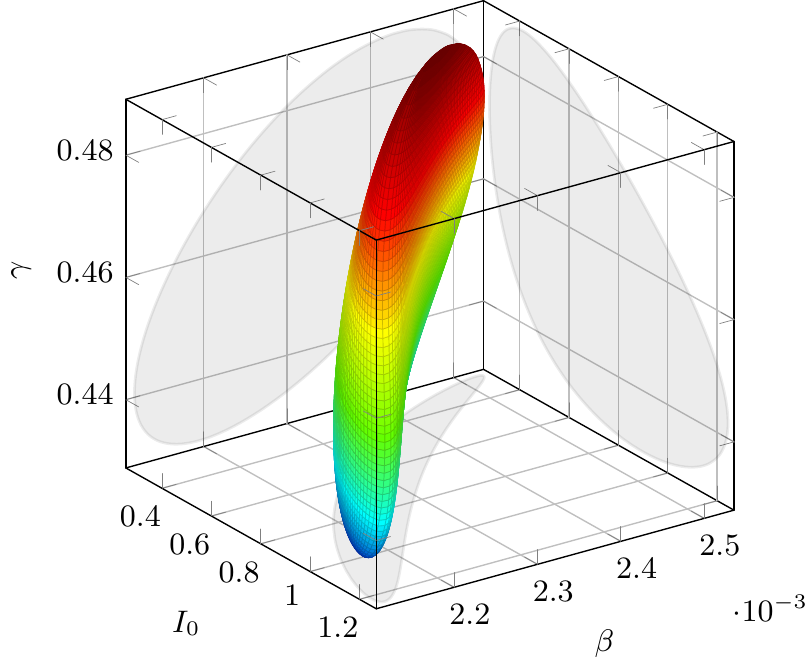}{%
		\begin{tikzpicture}
			\begin{axis}[view={55}{25}, xlabel={$I_0$}, ylabel={$\beta$}, zlabel={$\gamma$}, enlargelimits=false, legend style={font=\footnotesize}, small, width=.47\textwidth, height=.47\textwidth, colormap/bluered,
				3d box=complete, xmin=0.2533648, xmax=1.268026, ymin=0.002106662, ymax=0.002535686, zmin=0.428833, zmax=0.4891406,
				]
				\addplot3[thick, draw=black, fill=black, opacity=0.075, fill opacity=0.075] table [x index=0, y index=1, z expr ={0.428833}, col sep=comma] {PlotData/SIRshadow12.txt} --cycle;
				\addplot3[thick, draw=black, fill=black, opacity=0.075, fill opacity=0.075] table [x index=0, y expr={0.002535686}, z index=1, col sep=comma] {PlotData/SIRshadow13.txt} --cycle;
				\addplot3[thick, draw=black, fill=black, opacity=0.075, fill opacity=0.075] table [x expr={0.2533648}, y index=0, z index=1, col sep=comma] {PlotData/SIRshadow23.txt} --cycle;
				%
				\addplot3[patch, patch type=bilinear, shader=faceted, patch table={PlotData/Faces60.txt}, very thin] file {PlotData/Points60.txt};
			\end{axis}
	\end{tikzpicture}}
	\caption{Surface plot of 3D confidence boundary of level $1\sigma$ for the parameter triple $\theta = (I_0, \beta, \gamma)$ whose MLE is given by $\theta_\text{MLE} \approx (0.614, 0.00231, 0.458)$. Especially from the projected shadows, one can see that there is a slight bend in the confidence region as expected from the non-linear definition of the model.}
	\label{fig:SIR3D}
\end{figure}
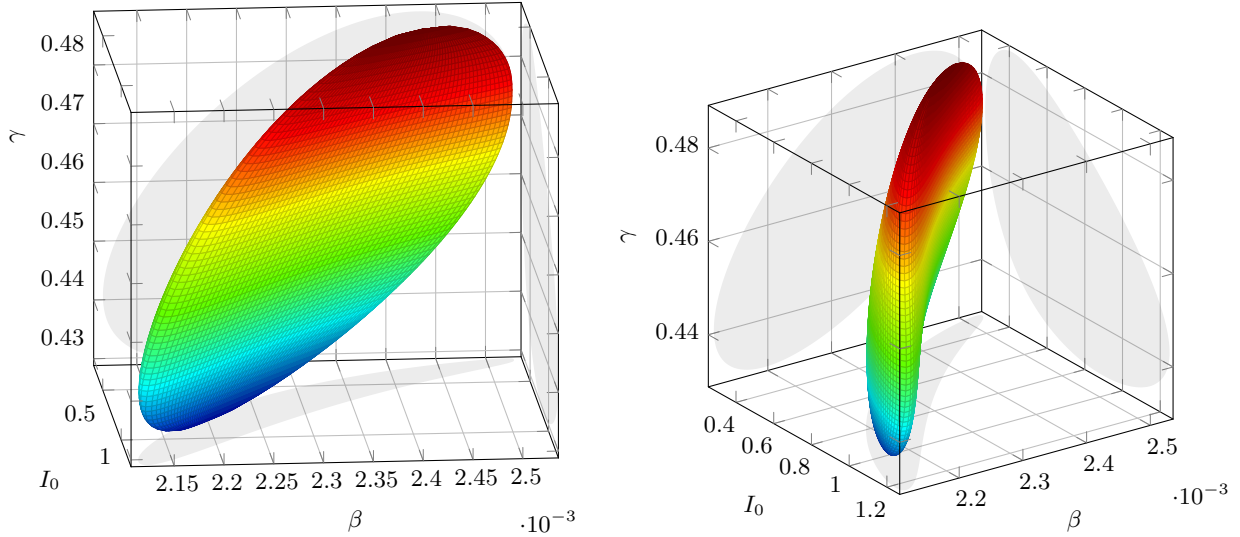

The parameter covariance matrix of confidence level $q = 1\sigma \approx \SI{68}{\%}$ is typically approximated as 
\begin{equation}
	\underbrace{F^{-1}_k(q)}_{\approx \num{3.53}} \cdot \, g^{-1}(\theta_\text{MLE}) \approx %
	 	\bqty{\begin{array}{r r r}
			\num{0.23}\hphantom{\cdot10^{-3}} & \num{-1.0e-4} & \num{-7.9e-3}\\
			\num{-1.0e-4} & \num{4.5e-8} & \num{3.9e-6}\\
			\num{-7.9e-3} & \num{3.9e-6} & \num{8.8e-4}\\
		\end{array}}
\end{equation}
with $\inv{F}_k$ the quantile function of the $\chi^2_k$-distribution, which scales the parameter covariance to the desired confidence level. On top of the synergistic effects which can be read off from the off-diagonal elements of this matrix, the exact confidence boundary visualised in \cref{fig:SIR3D} provides a more nuanced insight into the interdependent effects of the model parameters on the predictions. For instance, one can see from the amount of distortion in the respective projections that the pair-wise non-linear interactions are strongest between $I_0$ and $\beta$ and weakest between $\beta$ and $\gamma$.

%

Practitioners of the profile likelihood method will recognise the projected shadows of the confidence region as the higher-dimensional analogues of one-dimensional likelihood profiles: the parameters which have been projected out can be considered to have been set to their optimal values at every point in the projection. Thus the projections depicted in \cref{fig:SIR3D} respectively constitute the sets of 2D configurations for which the likelihood ratio does not exceed the $1\sigma$ threshold irrespective of the value of the remaining \enquote{nuisance} parameter. For models with $\dim \M > 3$, the same principle can be applied to study the parameter manifold, for instance by visualising three-dimensional slices at a time via projections of the high-dimensional exact confidence regions.

In contrast to projections, one can alternatively study conditionalisations of the model. For instance, one might fix the initial value to $I_0 \approx 0.61$ and thereby only explore the intersection of the $\beta$-$\gamma$ plane with the three-dimensional confidence region. By keeping the degrees of freedom fixed at three, it is possible to retain a one-to-one correspondence between the confidence boundaries determined in the 2D case of the $\beta$-$\gamma$ plane and the confidence boundaries in the 3D case.

\subsection{Performance and Complexity} \label{sec:Performance}

\FigureNames{Performance}


When it comes to constructing exact confidence regions, the main alternative to using the proposed integral manifold method is essentially given by sampling the log-likelihood on a (possibly non-uniform) grid of parameter configurations to determine between which grid vertices the confidence boundary of interest is located. For instance, this can be achieved using variants of the Marching Squares or Marching Cubes algorithms in 2D and 3D respectively \cite{MarchingCubes}. To pinpoint the intermediate crossing points with higher precision, piecewise polynomial approximations can also be used to interpolate between the sampled points. Although one could also use Monte Carlo simulation to produce a point cloud whose density is bijectively related to the value of the likelihood, iso-contours constructed from this cloud density are typically very irregular and imprecise. Moreover, this can become prohibitively expensive for higher confidence levels: as the point cloud thins out radially, the number of Monte Carlo samples must be increased to sustain useful precision in the estimate of the boundary locations.

The inherent disadvantage of such approaches is that the overwhelming majority of points where the log-likelihood is sampled are far away from the confidence boundary of interest, leading to a tremendous waste of computational resources. This is further exacerbated for higher-dimensional parameter manifolds and sampling grids. For this reason, investigations of exact confidence regions in practical applications have received little attention in the literature as they are typically considered to be computationally infeasible, particularly for large datasets and high-dimensional parameter manifolds.



It is straightforward to see that in the example of a two-dimensional globally structurally identifiable model, every confidence boundary is topologically equivalent to a circle around the MLE. To parametrise said boundary as an integral curve to a likelihood-annihilating vector field, the log-likelihood gradient $\dd \ell$ must be calculated at every point where said vector field is to be evaluated. A single calculation of the components of the log-likelihood gradient takes on the order of $\dim\M \cdot N$ steps for $N$ data points. Furthermore, assuming this needs to be repeated $H$ times along the topological circle, one ends up with $\dim\M \cdot H \cdot N$ evaluations overall. In comparison, the grid method for a two-dimensional parameter space takes on the order of $N$ steps per evaluation of the log-likelihood, which must be calculated on a grid of $H \cross H = H^{\dim \M}$ uniformly spaced points (although this may be a different value of $H$). Already, the overall complexity of the calculation is on the order of $N \cdot H^2$. In general, the grid sampling involved in the interpolation scheme requires on the order of $\Complexity{N \cdot H^{\dim\M}}$ evaluations of the log-likelihood whereas the integral manifold method only necessitates $\Complexity{N \cdot \dim\M \cdot H^{\dim\M-1}}$ evaluations due to having to integrate likelihood-annihilating vector fields along ${\dim\M-1}$ directions which suggests that the integral manifold method will generally outperform any grid sampling methods. 

In hindsight, this relationship in the scaling behaviours of both methods is unsurprising, given that Stokes' theorem $\int_U \dd{\omega} = \int_{\partial U} \, \omega$ reveals that the operation of taking the topological boundary of a set is intimately connected to the derivative operator. In other words, since the integral manifold method only samples the boundary of the confidence region, its scaling behaviour $\Complexity{\dim\M \cdot H^{\dim\M-1}}$ essentially corresponds to the derivative with respect to $H$ of the scaling behaviour $\Complexity{H^{\dim\M}}$ of the grid method, which samples the entire region.


\begin{figure}[!ht]
	\centering
	\begin{subfigure}[b]{0.43\textwidth}
		\centering
		\begin{small}
			\begin{tabular}{c c c} \toprule
				Solver & Function & Time \\
				Tolerance & Evaluations & \un{s} \\ \midrule
				$10^{-5}$  & \num{285}  & \num{0.40 \pm 0.01} \\
				$10^{-6}$  & \num{339}  & \num{0.47 \pm 0.01} \\
				$10^{-7}$  & \num{465}  & \num{0.62 \pm 0.01} \\
				$10^{-8}$  & \num{657}  & \num{0.88 \pm 0.03} \\
				$10^{-9}$  & \num{969}  & \num{1.32 \pm 0.03} \\
				$10^{-10}$ & \num{1329} & \num{1.79 \pm 0.05} \\
				$10^{-11}$ & \num{1797} & \num{2.46 \pm 0.17} \\
				$10^{-12}$ & \num{2709} & \num{3.82 \pm 0.24} \\
				$10^{-13}$ & \num{4317} & \num{6.21 \pm 0.32} \\
				$10^{-14}$ & \num{6837} & \num{9.82 \pm 0.67} \\
				\bottomrule
			\end{tabular}
		\end{small}
		\caption{Table of performance of confidence boundary generation scheme.}
		\label{subfig:PerformanceSCPtable}
	\end{subfigure}\hfill
	\begin{subfigure}[b]{0.53\textwidth}
		\centering
		\PublishGraphic[Saved]{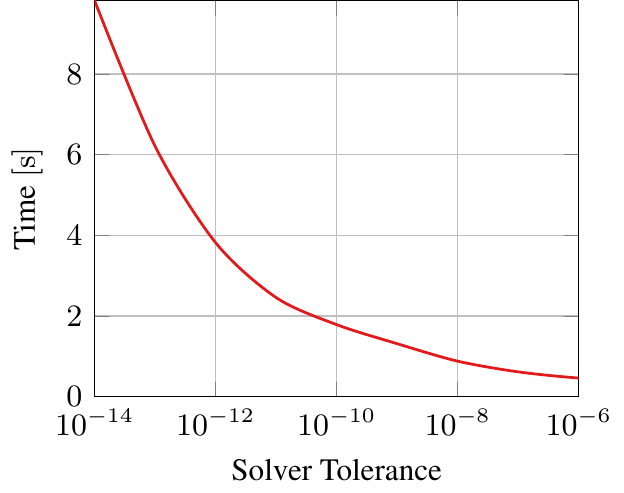}{\begin{tikzpicture}
			\begin{semilogxaxis}[ylabel={Time \un{s}}, xlabel={Solver Tolerance}, width = 1.1\textwidth, height=.75\textwidth, enlargelimits=false, xmin=1e-14, ymin=0, xmax=1e-6, small, xtick={1e-14, 1e-12, 1e-10, 1e-8, 1e-6}]
				\addplot+[smooth] coordinates {
				(1e-5, 0.3963)
				(1e-6, 0.465)
				(1e-7, 0.6184)
				(1e-8, 0.883)
				(1e-9, 1.316)
				(1e-10, 1.789)
				(1e-11, 2.46)
				(1e-12, 3.82)
				(1e-13, 6.21)
				(1e-14, 9.82)};
			\end{semilogxaxis}
		\end{tikzpicture}}
		\caption{Performance benchmark of the integral curve generation scheme in two dimensions.}
	\end{subfigure}
	\caption{Performance of the integral curve scheme for the determination of the $1\sigma$ confidence boundary on the SCP dataset, which contains $N=580$ observations. The evaluation of the log-likelihood for this dataset was measured as \SI{0.64}{ms} while its gradient was measured as taking \SI{0.87}{ms} on average. All calculations were executed in a single core computation and using the Tsitouras $5^\text{th}$ order Runge--Kutta algorithm \cite{Tsitouras}. Further details and system specifics can be found in \secref[sec:PerformanceAndComplexityDetails]{appendix \ref*{sec:PerformanceAndComplexityDetails}}.}
	\label{fig:PerformanceSCP}
\end{figure}
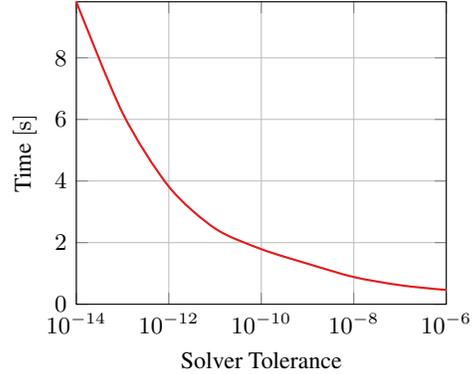

\Cref{fig:PerformanceSCP} indicates the performance of the integral manifold method for the SCP dataset from \cref{sec:SCPAnalysis}. Although time measurements are specific to the system on which the benchmarks were executed, the number of required function evaluations per solve provides a deterministic system-independent performance measure.

\Cref{subfig:PerformanceSCPtable} lists the number of evaluations of the log-likelihood gradient which are required in the numerical integration of the ODE from \cref{eqn:IntegralCurve} in order to obtain a closed integral curve, given an initial point that is already known to lie exactly on the confidence boundary. That is, the cited numbers of function evaluations exclude the process of locating the initial point on the confidence boundary of interest. However, given that such a point can always be located via a one-dimensional search on a radial line emanating from the MLE (e.g.\ using the Newton-Raphson method or bisection), the computational effort required is generally insignificant compared with the subsequent ODE integration. As illustrated by \cref{subfig:PerformanceSCPtable}, less than $500$ evaluations of the log-likelihood gradient can be sufficient to locate the confidence boundary to within a relative tolerance of $10^{-7}$.

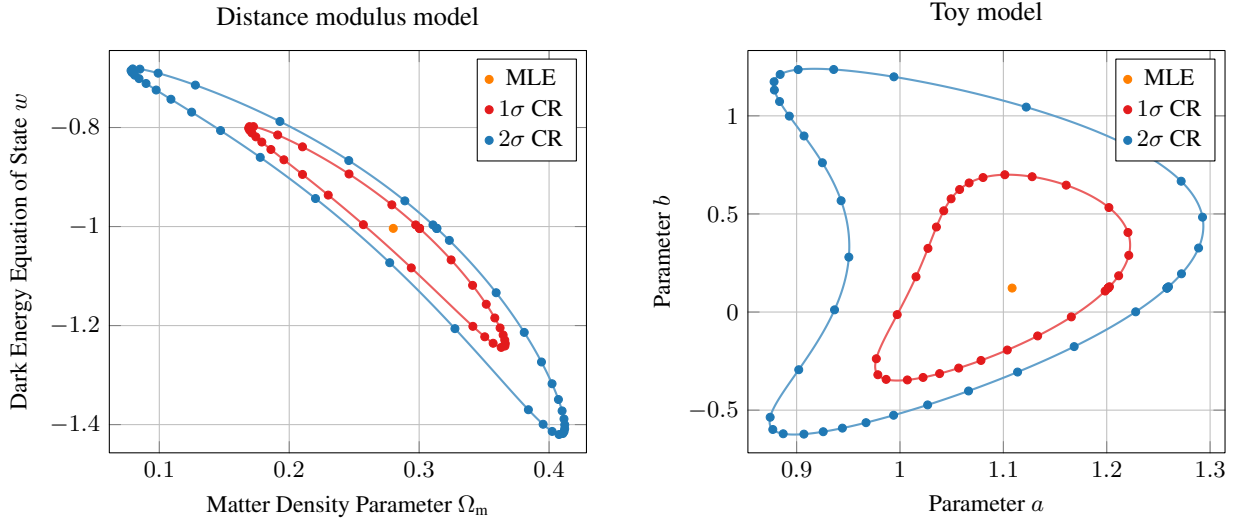
\begin{figure}[!ht]
	\centering
	\PublishGraphic[Saved]{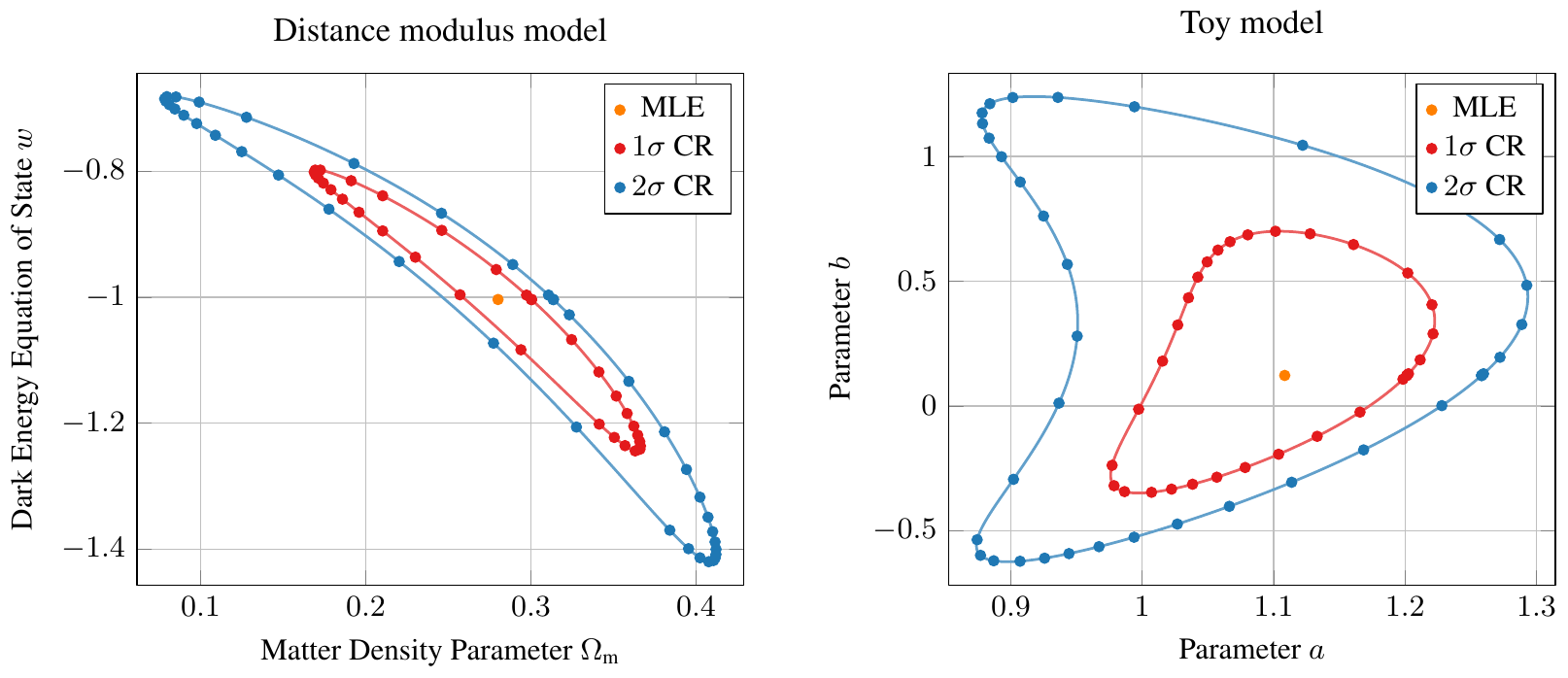}{
		\begin{tikzpicture}
			\pgfplotstableread[col sep=comma]{PlotData/SCPConf1sigma.txt}{\ConfONE}
			\pgfplotstableread[col sep=comma]{PlotData/SCPConf2sigma.txt}{\ConfTWO}
			\pgfplotstableread[col sep=comma]{PlotData/SCPSol1Points.txt}{\PointsONESCP}
			\pgfplotstableread[col sep=comma]{PlotData/SCPSol2Points.txt}{\PointsTWOSCP}
			\pgfplotstableread[col sep=comma]{PlotData/ModelTiles/model10.txt}{\Conf}
			\pgfplotstableread[col sep=comma]{PlotData/ToySol1Points.txt}{\PointsONE}
			\pgfplotstableread[col sep=comma]{PlotData/ToySol2Points.txt}{\PointsTWO}
			\begin{groupplot}[group style={group size=2 by 1, horizontal sep=0.13\textwidth}, legend style={font=\footnotesize,text=black}]
			\nextgroupplot[title={Distance modulus model}, xlabel={Matter Density Parameter $\Omega_\text{m}$}, ylabel={Dark Energy Equation of State $w$}, enlarge x limits=0.05, small, width=.48\textwidth, height=.42\textwidth, xtick distance = 0.1, ytick distance = 0.2,
			legend entries={MLE, $1\sigma$ CR, $2\sigma$ CR}]
				\addplot+[smooth,draw=mycol1!70!white,forget plot] table [x index = 0, y index=1] {\ConfONE};
				\addplot+[smooth,draw=mycol2!70!white,forget plot] table [x index = 0, y index=1] {\ConfTWO};
				\addplot[only marks, mark size=1.5pt, fitcol] coordinates {(0.28007,-1.00369)};
				\addplot[only marks, mycol1, mark size=1.5pt] table [x index = 0, y index=1] {\PointsONESCP};
				\addplot[only marks, mycol2, mark size=1.5pt] table [x index = 0, y index=1] {\PointsTWOSCP};
			\nextgroupplot[title={Toy model}, xlabel={Parameter $a$}, ylabel={Parameter $b$}, enlarge x limits=0.05, small, width=.48\textwidth, height=.42\textwidth, xtick distance = 0.1, ytick distance = 0.5,
			legend entries={MLE, $1\sigma$ CR, $2\sigma$ CR}]
				\addplot+[smooth,draw=mycol1!70!white,forget plot] table [x index = 0, y index=1] {\Conf};
				\addplot+[smooth,draw=mycol2!70!white,forget plot] table [x index = 2, y index=3] {\Conf};
				\addplot[only marks, mark size=1.5pt, fitcol] coordinates {(1.10847, 0.122159)};
				\addplot[only marks, mycol1, mark size=1.5pt] table [x index=0, y index=1] {\PointsONE};
				\addplot[only marks, mycol2, mark size=1.5pt] table [x index=0, y index=1] {\PointsTWO};
			\end{groupplot}
		\end{tikzpicture}
	}
	\caption{Visualisation of base points of the ODE solutions to the confidence boundary integral curves computed using the Tsitouras $5^\text{th}$ order Runge--Kutta algorithm, together with its associated free $4^\text{th}$ order interpolation \cite{Tsitouras}. Left-hand side: SCP model from \cref{eqn:DistModulus}, right-hand side: non-linearly parametrised toy model $y_\model(x; a,b) = (a+b) \, x + \exp(a-b)$ from \cref{sec:Survey}. The displayed ODE solutions were computed to a relative tolerance of $10^{-5}$. This illustrates the added efficiency of the integral manifold scheme that is derived from the use of adaptive ODE solvers, namely by spending less computational resources in regions where the curvature of the ODE solution is low.}
	\label{fig:SolPoints}
\end{figure}

\Cref{fig:SolPoints} exemplifies confidence boundaries obtained as solutions to the numerical integration of likelihood-annihilating vector fields together with their base points, at which said vector fields were evaluated. 
The deviation between the starting point of the integration and its termination after one full revolution around the MLE provides a measure of the accumulated global truncation error in the numerical integration and accordingly can be used a-posteriori to confirm that the obtained solution indeed conforms to the specified tolerance.





\section{Discussion}


Exact simultaneous confidence regions not only provide an accurate reflection of the uncertainty associated with the best fit parameters of a model, but moreover allow for nuanced insight into the 
structure of a model by faithfully illustrating non-linear interdependencies of its parameters. 
Above all, precise quantifications of parameter uncertainties are required for a meaningful propagation of the parameter uncertainty to the model predictions which arguably constitutes the most important part of the inference process.

The substantial computational effort involved in locating exact confidence regions, which results from the need to evaluate the log-likelihood for different parameter configurations, has lead many researchers to routinely rely on imprecise approximations of the parameter uncertainties instead. 
In this work, we showed how the differentiability of structurally identifiable models can be exploited to significantly reduce this computational cost, making precise parameter uncertainty analyses feasible for a wider class of problems.



First, we reviewed the definition of the Fisher metric via the Hessian of the Kullback--Leibler divergence in \cref{sec:DivergencesAndFisherMetric}. In particular, we noted that for Gaussian observations, the inverse covariance matrix $\inv{\Sigma}$ constitutes an inner product on the data space $\Y^N$. Moreover, the resulting pull-back metric $g_\M = h^* g_{\Y^N}$ induced on the parameter manifold coincides with the Fisher metric as prescribed via the Hessian of the Kullback--Leibler divergence. 
%
%
%
Next, the conventional definition of confidence regions via the likelihood ratio test was summarised and its assumption of the large sample limit through Wilks' theorem discussed in \cref{sec:DefiningConfidenceRegions}. 



\Cref{sec:ParameterIdentifiability} summarised some of the established terminology concerning the topic of parameter identifiability. In particular, we highlighted the fact that structural identifiability essentially signifies the injectivity of a model map with respect to its parameters on various domains, e.g.\ in local neighbourhoods around a point or globally on \M. Moreover, a parametrisation-invariant criterion for the assessment of local structural identifiability is given by the determinant of the Fisher metric $\det(g)$, which vanishes in the presence of local structural non-identifiabilities.


In \cref{sec:IsoLikelihoodSurfaces}, we derived a systematic approach for constructing a set of vector fields whose integral manifolds constitute the boundaries of confidence regions. Therefore, the problem of locating confidence boundaries is converted to solving a system of ODEs given an initial point which is already known to lie on the boundary of interest and which can be determined via a one-dimensional line search in the parameter space. Moreover, we prove in \secref[sec:LieAlgebraClosure]{appendix \ref*{sec:LieAlgebraClosure}} that on subregions of the parameter manifold \M\ where the model is injective, the thus constructed set of vector fields forms a closed $(\dim \M -1)$-dimensional Lie algebra. Therefore, the confidence boundaries are guaranteed to foliate this subregion of the parameter manifold by virtue of Frobenius' theorem.

%
%
%
%

In \cref{sec:ConfidenceBands} we demonstrated how pointwise confidence bands can be constructed around the predictions of a model from the exact confidence region. Given the definition of likelihood-based confidence regions, the resulting confidence bands constitute a faithful probabilistic assessment of the uncertainties in the model predictions, predicated on the assumption that the given model is correct, i.e.\ that the observed data has indeed been generated by the given model. Further, \secref[sec:TopologicalProof]{appendix \ref*{sec:TopologicalProof}} proves that if a given model is injective with respect to its parameters on the closure $\overline{\mathcal{C}_q}$ of a bounded confidence region $\mathcal{C}_q$, the confidence bands only require the model to be evaluated on the boundary $\partial \mathcal{C}_q$ of a confidence region but not in its interior, which further reduces the computational effort involved. Although visualisations of confidence boundaries remain practically limited to three-dimensional slices of the parameter manifold at a time, the full high-dimensional surfaces can nevertheless be used in computations, e.g.\ for uncertainty propagation in the form of pointwise confidence bands.





\Cref{sec:Survey} illustrated the effects of non-linear model reparametrisations on confidence regions for a given toy dataset. It was noted that such reparametrisations affect not only the location and size of said confidence regions but can also strongly distort their shapes. Given that the likelihood approaches a normal distribution in the asymptotic limit of $N \longrightarrow \infty$, this distortion is typically more pronounced for small datasets. Also, the magnitude of this non-linear coordinate distortion on the parameter space usually increases with radial distance from the MLE. In aggregate, this effects that for non-linearly parametrised models, confidence regions must be computed separately for every confidence level of interest since confidence regions of differing levels are no longer of similar geometric shape.

\Cref{sec:Performance} compared the computational complexity of the integral manifold method as a function of the dimensionality of the parameter space \M\ against grid sampling methods for the construction of exact confidence regions. Whereas grid sampling approaches exhibit an algorithmic complexity of $\Complexity{H^{\dim\M}}$, it was found that the integral manifold method generally scales according to $\Complexity{\dim\M \cdot H^{\dim \M -1}}$. In addition, the integral manifold method directly benefits from the use of adaptive ODE solvers. Specifically, for the real-world example of the cosmological distance modulus model from \cref{eqn:DistModulus} with $\dim \M=2$, it was observed that less than $500$ evaluations of the log-likelihood gradient are sufficient to locate the $1\sigma$ boundary to within a relative tolerance of $10^{-7}$. The fact that this integration results in closed curves is a testament to the reliability of this method.


%



Another approximative scheme that attempts to improve upon the ellipsoidal approximation of confidence regions is given by the Derivative Approximation for Likelihoods (DALI) \cite{Sellentin1, Sellentin2} which relies on expansions of the log-likelihood in a Taylor series with respect to the model parameters. As a result, one only has rather coarse-grained control over the accuracy of DALI approximations through the choice of the order up to which the likelihood is expanded. Moreover, there is no straightforward way to quantify the accuracy of a given DALI approximation for the purpose of determining confidence regions without also computing the exact result using the full likelihood and performing a side-by-side comparison. The accuracy of the DALI approximation degrades with increasing radius of the confidence region, i.e.\ for higher confidence levels. In contrast, the proposed integral manifold method allows for more fine-grained control of the accuracy by specifying relative and absolute tolerance to the solver algorithm in the integration of the ODE system. Furthermore, the precision with which the confidence regions are located and the computational cost of their construction are essentially independent of the confidence level.



Another popular device for investigating parameter uncertainties in non-linear models is the so-called profile likelihood method, in which the parameter space is explored on one-dimensional curves emanating radially from the MLE \cite{Timmer}. This one-dimensional sampling intentionally ignores interactive effects between the parameters and instead mainly focuses on placing bounds on their values individually. Although the profile likelihood method can be extended to higher-dimensions such that pair-wise interactions of parameters are taken into account, this again runs into the aforementioned problem of computational complexity due to the requirement of evaluating the log-likelihood on higher-dimensional grids and thus spending valuable computational resources far away from the confidence boundary of interest \cite{HeldAppliedStatisticalInference, Sprott}.



The bulk of the computational effort associated with the profile likelihood scheme results from the need to reoptimise all \enquote{nuisance} parameters at every step, which is not required by the integral manifold method. %
%
%
%
On the other hand, whereas the integral manifold scheme requires structurally non-identifiable parameter combinations to be eliminated from a given model before it can be meaningfully employed, the profile likelihood method can be applied to models irrespective of their non-identifiability. Moreover, for particularly high-dimensional parameter manifolds, independent parameter uncertainty analyses in terms one-dimensional likelihood profiles can be more straightforward in terms of their interpretation. In summary, this makes the profile likelihood method a robust fallback for scenarios to which the presented integral manifold scheme is not suited.


Next, we illustrated the benefits of analysing exact confidence regions instead of their approximations in real-world examples from fundamental physics and systems biology. \Cref{sec:SCPAnalysis} discussed the cosmological distance modulus model, which relates the apparent distance of type Ia supernov\ae\ to their redshift under the assumption of a flat Universe. Specifically, this model was applied to a dataset recorded by the Supernova Cosmology Project, which consists of 580 observations. Here, the integral manifold method not only exhibited great performance but also allowed for a precise quantification of the uncertainty in the MLE which is especially desirable since the parameters of the distance modulus model directly correspond to fundamental cosmological constants. The confidence bands associated with the maximum likelihood prediction indicated that further measurements at high redshifts might serve best to further constrain the parameters, which is consistent with the underlying physical theory.




As a real-world example for applications in systems biology, where mathematical models are often formulated in terms of differential equations, we illustrated the use integral manifold method via the topical class of SIR models in \cref{sec:DiseaseModelling}. Although we selected the structurally simplest member of this family for the purpose of demonstration, many of today's state of the art methods for predicting the spread of infectious diseases such as COVID-19 are nevertheless direct logical descendants of this model \cite{SpoCK, RackauckasCovid}.

Due to its non-monotonic nature, the SIR model portrayed the merits of exact confidence bands well, which exhibited a complex and asymmetric structure around the maximum likelihood prediction. Notably, the prediction uncertainty was found to be largest around the peak of the infection wave, which indicates that further observations at said peak would be most effective in constraining both the model parameters as well as subsequent predictions. Intuitively, this can be explained by the fact that the location of the peak is collectively influenced by all three parameters of the SIR model which means that there is a compounding effect of the collective uncertainties in the parameter values on the predictions in this part of the time domain.

%
Both the distance modulus and SIR models are simple enough to make experimental design deductions directly from their respective mechanistic structures. However, there are many instances of dynamical models which exhibit more complex behaviour, such that it is no longer feasible to plan experiments based on theoretical reasoning. This is precisely where confidence bands can provide a useful tool for experimental design, as they depict the collective influence of the parameter uncertainties on the predictions in an accessible form.


For the cosmological distance modulus model, the main source of computational effort in computing the log-likelihood derives from the large number of data points on which the model must be evaluated to generate predictions. In contrast, the SIR model was applied to a dataset containing only fourteen observations. However, the ODE system underlying the SIR model must be numerically integrated to within a specified tolerance every time a set of predictions is generated for a different parameter configuration.




In summary, the main benefits derived from use of the integral manifold method can differ depending on the context: small data applications such as the systems biological SIR example profit chiefly from the detailed consideration of non-linear distortions of confidence regions which are typically more pronounced due to the low number of observations. For large data applications, the shape of the likelihood is usually closer to a Gaussian distribution. However, due to the increased number of observations for which the model predictions must be computed, the likelihood becomes costlier to evaluate. Therefore, while the non-linear distortion in the confidence regions may be lower for large data settings, the main advantage of the integral manifold method stems from its economical evaluations of the likelihood and its derivatives.

We also provide an open source implementation of the integral manifold scheme and other methods via the \InfoGeo\ package \cite{InformationGeometryJL} for the \href{https://julialang.org/downloads/}{Julia programming language}. The supplementary material includes the source code required to define all models discussed in this work which allows for convenient reproduction of the presented results \cite{SupplementaryCode}. Further examples of how to use \InfoGeo\ can be found in the \href{https://rafaelarutjunjan.github.io/InformationGeometry.jl/dev/}{associated documentation}.

We limited the scope of the discussion to datasets which only feature uncertainties in the dependent variables (i.e.\ the $y$-values) of a dataset. We aim to address the construction of exact confidence regions for more general datasets with mixed uncertainties in both the dependent and independent variables (i.e.\ both the $y$ and $x$-variables) in future work.


\FigureNames{Appendix}

\section*{Appendix}
\subsection{Closure of Lie Algebra of Likelihood-Annihilating Vector Fields} \label{sec:LieAlgebraClosure}

Frobenius' theorem guarantees that the span of a set of vector fields $X_1,...,X_k \in \VF{\M}$ generates a unique family of integral manifolds if and only if said span constitutes a closed Lie subalgebra of $\VF{\M}$ \cite{Lee}. If this family of integral manifolds indeed exists, it is also guaranteed to foliate \M. This section aims to investigate whether vector fields of the form given in \cref{eqn:OrthVF} constitute a closed Lie algebra. Specifically, the set of smooth vector fields of this form will be denoted by
\begin{equation}
	\mathfrak{L} \coloneqq \,\Setcond{K(\vec{\alpha}) \in \VF{\M}}{\vec{\alpha} \in \mathcal{H}} = K(\mathcal{H})
\end{equation}
from here on out, where $K(\vec{\alpha})$ denotes the collection of $K_p(\vec{\alpha})$ for all $p \in \M$. By definition, one therefore has $\Lie{X} \ell = 0$ for all $X \in \mathfrak{L}$, meaning that any element of $\mathfrak{L}$ annihilates the log-likelihood.


The proof outlined in this section highlights that the necessary restrictions consist of the local structural identifiability of the model on the one hand and twice-continuous differentiability of the log-likelihood $\ell$. However, for sake of notational simplicity, we will assume that the likelihood is smooth with respect to the parameters in the following discussion.

It is not hard to see that the set $\mathfrak{L}$ must be smaller than the set $\,\Setcond{X \in \VF{\M}}{\Lie{X} \ell \equiv X \ell= 0}$ since not all vector fields which annihilate $\ell$ are necessarily of the form given in \cref{eqn:OrthVF}. That is, if $X \in \mathfrak{L}$, then any other smooth vector field $Y$ which is related to $X$ by a smooth function provides another valid solution to \cref{eqn:ScoreAnnihilation}, i.e.\
\begin{equation}
	\forall X \in \mathfrak{L} : \forall f \in \Smooth{\M} : \qquad \qquad Y = f X \qquad \Longrightarrow \qquad \Lie{Y} \ell = 0
\end{equation}
while generally $Y \notin \mathfrak{L}$. Using the vector space isomorphism $K_p : \R^{\dim\M} \longrightarrow T_p\M$, it immediately follows that $\pqty{\mathfrak{L},+,\cdot\,}$ forms a (finite-dimensional) $\R$-vector subspace of $\pqty{\VF{\M},+,\cdot\,}$ since $\mathcal{H}$ is an $\R$-vector subspace of $\R^{\dim\M}$. Thus, it only remains to be shown that $\mathfrak{L}$ is closed with respect to the Lie bracket, i.e.\ that $[X,Y] \in \mathfrak{L}$ for all $X,Y \in \mathfrak{L}$. To show this, it is again convenient to make use of the isomorphism $K_p$. The vector space $\mathcal{H}$ can be equipped with a Lie bracket $\llbracket \openslot, \openslot \rrbracket : \mathcal{H} \cross \mathcal{H} \longrightarrow \mathcal{H}$ in such a way that it is compatible with the Lie bracket of smooth vector fields in the sense
\begin{equation}
	K_p\pqty{\llbracket \vec{\alpha}, \vec{\beta} \rrbracket} \overset{!}{=} \bqty{K_p(\vec{\alpha}),K_p(\vec{\beta})}.
\end{equation}
Clearly, this condition is satisfied by just using $K_p$ to define the bracket as
\begin{equation}
	\llbracket \vec{\alpha}, \vec{\beta} \rrbracket \coloneqq K_p^{-1}\pqty{\bqty{K_p(\vec{\alpha}),K_p(\vec{\beta})}}
\end{equation}
since $K_p$ is invertible. From this definition, it follows that $\mathcal{H}$ and $\mathfrak{L}$ must be isomorphic as Lie algebras provided that they are both closed under their respective Lie brackets, which can be summarised as
\begin{equation}
	\pqty{\mathcal{H},\llbracket \openslot, \openslot \rrbracket} \cong_\text{Lie alg.} \pqty{\mathfrak{L},[\openslot,\openslot]}
\end{equation}
because the isomorphism $K_p$ is valid at every point $p \in \M$. The problem of proving that $\mathfrak{L}$ is closed with respect to the Lie bracket $[\openslot,\openslot]$ is thus reduced to showing that $\mathcal{H}$ is closed in $\R^{\dim\M}$ with respect to the Lie bracket $\llbracket \openslot, \openslot \rrbracket$.

For any smooth vector fields $X,Y \in \VF{\M}$ one can express the Lie bracket in components as
\begingroup\allowdisplaybreaks
\begin{align}
	[X,Y]f &= X(Yf) - Y(Xf) = X^i \, \pdv{\theta^i}\pqty{ Y^j \, \pdv{f}{\theta^j} } - Y^i \, \pdv{\theta^i}\pqty{ X^j \, \pdv{f}{\theta^j} } \\
	&= X^i \, \pqty{\pdv{Y^j}{\theta^i} \, \pdv{f}{\theta^j} + Y^j \pdv{f}{\theta^i\,}{\theta^j}} - Y^i \, \pqty{\pdv{X^j}{\theta^i} \, \pdv{f}{\theta^j} + X^j \pdv{f}{\theta^i\,}{\theta^j}} \\
	&= \pqty\bigg{\underbrace{X^i \, \pdv{Y^j}{\theta^i} - Y^i \, \pdv{X^j}{\theta^i}}_{=[X,Y]^j}} \pdv{f}{\theta^j} + \underbrace{(X^i \, Y^j - Y^i \, X^j) \, \pdv{f}{\theta^i\,}{\theta^j}}_{=0}
\end{align}\endgroup%
where the last term vanishes due to the contraction of a symmetric with an antisymmetric quantity. By representing the linear transformation $K_p$ via $\pqty\big{K_p(\vec{\alpha})}^{\! i} = \tensor{M}{^i _j} \, \alpha^j$, one can compute
\begingroup\allowdisplaybreaks
\begin{align}
	\pqty{K_p^{-1}\pqty{\bqty{K_p(\vec{\alpha}),K_p(\vec{\beta})}}}^{\!\!i} {}&{}= \tensor{(M^{-1})}{^i _j} \, \pqty\Big{\underbrace{\tensor{M}{^b _a} \, \alpha^a}_{[K_p(\vec{\alpha})]^b  \vphantom{[K_p(\vec{\beta})]^j}}  \, \partial_b \, \underbrace{\tensor{M}{^j _c} \, \beta^c}_{[K_p(\vec{\beta})]^j} - \tensor{M}{^b _a} \, \beta^a \, \partial_b \, \tensor{M}{^j _c} \, \alpha^c}\\
	{}&{}= \pqty{\alpha^a \, \beta^c - \alpha^c \, \beta^a} \, \tensor{(M^{-1})}{^i _j} \, \tensor{M}{^b _a} \, \partial_b \, \tensor{M}{^j _c}
\end{align}\endgroup%
where the abbreviation $\partial_b \coloneqq \pdv*{\smash{\theta^b}}$ was used. Further, by use of the chain rule one finds
\begin{equation}
	0 = \pdv{\theta^b} \pqty{\delta^i_c} = \pdv{\theta^b} \pqty{\tensor{(M^{-1})}{^i _j} \, \tensor{M}{^j _c}} = \pdv{\tensor{(M^{-1})}{^i _j}}{\theta^b} \, \tensor{M}{^j _c} + \tensor{(M^{-1})}{^i _j} \, \pdv{\tensor{M}{^j _c}}{\theta^b}
\end{equation}
from which it immediately follows that $\tensor{(M^{-1})}{^i _j} \, \partial_b \, \tensor{M}{^j _c} = -\tensor{M}{^j _c} \, \partial_b \, \tensor{(M^{-1})}{^i _j}$, i.e.\ the derivative can be shifted from the matrix $M$ onto its inverse $M^{-1}$ at the cost of a minus sign.
\begin{align}
	\llbracket \vec{\alpha}, \vec{\beta} \rrbracket^i = \bqty{K_p^{-1}\pqty{\bqty{K_p(\vec{\alpha}),K_p(\vec{\beta})}}}^{\! i} {}&{}= \pqty{\alpha^a \, \beta^c - \alpha^c \, \beta^a} \, \tensor{(M^{-1})}{^i _j} \, \tensor{M}{^b _a} \, \partial_b \, \tensor{M}{^j _c}\\
	{}&{}= -\pqty{\alpha^a \, \beta^c - \alpha^c \, \beta^a} \, \tensor{M}{^j _c} \, \tensor{M}{^b _a} \, \partial_b \, \tensor{(M^{-1})}{^i _j}
\end{align}
The partial derivatives of the coefficient functions of $M^{-1}$ can be worked out as
\begin{align}
	\pdv{\tensor{(M^{-1})}{^i _j}}{\theta^b} &= \pdv{\theta^b} \bqty\Bigg{\frac{1}{B} \, \mathrm{diag}\pqty{\! \pqty{\pdv{\ell}{\theta^1}}, ..., \pqty{\pdv{\ell}{\theta^n}}\!}^{\!\!i}_{\,j}} = \pdv{\theta^b} \bqty{\frac{1}{B} \, \delta^i_j \, \pdv{\ell}{\theta^j}}\\
	&= -\frac{1}{B^2} \, \pdv{B}{\theta^b} \, \delta^i_j \, \pdv{\ell}{\theta^j} + \frac{1}{B} \, \delta^i_j \, \pdv{\ell}{\theta^b \,}{\theta^j} = \frac{1}{B} \, \delta^i_j \, \pqty{\pdv{\ell}{\theta^b \,}{\theta^j} - \pdv{\ell}{\theta^j} \, \pdv{\ln(B)}{\theta^b}}. \label{eqn:LieBracketResult}
\end{align}
Reinserting this expression for the partial derivatives of $M^{-1}$ yields
\begingroup\allowdisplaybreaks
\begin{align}
	\llbracket \vec{\alpha}, \vec{\beta} \rrbracket^i 
	& = -\pqty{\alpha^a \, \beta^c - \alpha^c \, \beta^a} \, \tensor{M}{^j _c} \, \tensor{M}{^b _a} \, \partial_b \, \tensor{(M^{-1})}{^i _j} \\
	&= -\pqty{\alpha^a \, \beta^c - \alpha^c \, \beta^a} \, \tensor{M}{^j _c} \, \tensor{M}{^b _a} \, \frac{1}{B} \, \delta^i_j \, \pqty{\pdv{\ell}{\theta^b \,}{\theta^j} - \pdv{\ell}{\theta^j} \, \pdv{\ln(B)}{\theta^b}}\\
	&= -2 \underbrace{\tensor{\pqty\big{K_p(\vec{\alpha})}}{^{\![b\vphantom{j}}} \, \tensor{\pqty\big{K_p(\smash{\vec{\beta})}}}{^{\!j]}} \vphantom{\pdv{\ell}{\theta^b \,}{\theta^j}}}_{\text{antisymm.}} \, \frac{1}{B} \, \delta^i_j \, \pqty\bigg{\underbrace{\pdv{\ell}{\theta^b \,}{\theta^j}}_{\text{symm.} \vphantom{\text{antisymm.}}} - \,\pdv{\ell}{\theta^j} \, \pdv{\ln(B)}{\theta^b}} \label{eqn:AntiSymmetriser1}\\
	&= 2 \tensor{\pqty\big{K_p(\vec{\alpha})}}{^{\![b\vphantom{j}}} \, \tensor{\pqty\big{K_p(\smash{\vec{\beta})}}}{^{\!j]}} \, \frac{1}{B} \, \delta^i_j \, \pdv{\ell}{\theta^j} \, \pdv{\ln(B)}{\theta^b} \label{eqn:AntiSymmetriser2}
\end{align}\endgroup%
where the sum over the index $j$ is inhibited by the Kronecker symbol $\delta^i_j$ which restricts the sum to the term corresponding to the open index $i$. Additionally, the expressions in \cref{eqn:AntiSymmetriser1,eqn:AntiSymmetriser2} employ the commonly used antisymmetrisation bracket notation. Since the above expression is an element of the vector space $\R^{\dim\M}$, it remains to be shown that $\vec{n} \dotp \llbracket \vec{\alpha}, \vec{\beta} \rrbracket = 0$ in order to guarantee that $\llbracket \vec{\alpha}, \vec{\beta} \rrbracket \in \mathcal{H}$. Finally, one obtains
\begingroup\allowdisplaybreaks
\begin{IEEEeqnarray}{rl}
	{}&{} \vec{n} \dotp \llbracket \vec{\alpha}, \vec{\beta} \rrbracket = \sum_{i=1}^{\dim\M} (1,...,1)^i \cdot \llbracket \vec{\alpha}, \vec{\beta} \rrbracket^i = \sum_{i=1}^{\dim\M} \llbracket \vec{\alpha}, \vec{\beta} \rrbracket^i \\
	={}&{} \sum_{i=1}^{\dim\M} 2 \tensor{\pqty\big{K_p(\vec{\alpha})}}{^{\![b\vphantom{j}}} \, \tensor{\pqty\big{K_p(\smash{\vec{\beta})}}}{^{\!j]}} \, \frac{1}{B} \, \delta^i_j \, \pdv{\ell}{\theta^j} \, \pdv{\ln(B)}{\theta^b}\\
	={}&{} \frac{1}{B} \, \pdv{\ln(B)}{\theta^b} \, \bqty{\pqty\big{K_p(\vec{\alpha})}^{\! b} \, \sum_{i=1}^{\dim\M} \delta^i_j \, \pqty\big{K_p(\vec{\beta})}^{\!j} \, \pdv{\ell}{\theta^j} - \pqty\big{K_p(\vec{\beta})}^{\! b} \, \sum_{i=1}^{\dim\M} \delta^i_j \, \pqty\big{K_p(\vec{\alpha})}^{\!j} \, \pdv{\ell}{\theta^j}} \IEEEeqnarraynumspace\\
	={}&{} \frac{1}{B} \, \pdv{\ln(B)}{\theta^b} \, \bqty\Bigg{\pqty\big{K_p(\vec{\alpha})}^{\! b} \, \underbrace{\sum_{j=1}^{\dim\M} \pqty\big{K_p(\vec{\beta})}^{\!j} \, \pdv{\ell}{\theta^j}}_{=0} - \,\pqty\big{K_p(\vec{\beta})}^{\! b} \, \underbrace{\sum_{j=1}^{\dim\M} \pqty\big{K_p(\vec{\alpha})}^{\!j} \, \pdv{\ell}{\theta^j}}_{=0}} = 0
\end{IEEEeqnarray}\endgroup%
where the summation over $j$ is now executed without obstruction. This causes the expression to vanish due to the contraction of the components of the vector field $K_p(\vec{\alpha})$ with the derivatives of the log-likelihood which vanishes by construction for any $\vec{\alpha} \in \mathcal{H}$. Thus, the new element $\vec{\nu} = \llbracket \vec{\alpha}, \vec{\beta} \rrbracket$ must be in $\mathcal{H}$ which then concludes the proof that $\mathfrak{L}$ is a closed Lie subalgebra of $\pqty{\VF{\M},+,\cdot\,,[\openslot,\openslot]}$.

Since vector fields constructed via \cref{eqn:OrthVF} evidently form a $(\dim\M -1)$-dimensional Lie subalgebra of the infinite-dimensional Lie algebra of smooth vector fields, Frobenius' theorem guarantees that integral manifolds of this subalgebra always exist. Moreover, the outlined proof identifies the sufficient differentiability of $\ell$ as well as the structural identifiability of the model as the key criteria for the guaranteed existence of confidence regions. 

Therefore, in the case of higher-dimensional confidence boundaries such as surfaces or manifolds in general, one can use the flows with respect to a basis of the Lie algebra $\mathfrak{L}$ to reach any point belonging to a connected confidence boundary from a given starting point on said boundary. Intuitively, this can also be imagined as meshing the confidence boundary using families of integral curves whose tangent vectors collectively form a $(\dim\M-1)$-dimensional linear subspace of the tangent space $T_p\M$ at every point $p$ on the confidence boundary.


\subsection{Evaluation of Models on the Confidence Boundary} \label{sec:TopologicalProof}

The topological relation to be proven for a continuous map $f:\M \longrightarrow \mathcal{Z}$ is given by
\begin{equation}
	\partial f(C) \subseteq f(\partial C)
\end{equation}
for some set $C \subseteq \M$. First, it is necessary to recall that a map $f$ between topological spaces $(\M,\mathcal{O}_\M)$ and $(\mathcal{Z},\mathcal{O}_{\mathcal{Z}})$ is said to be closed if it always maps closed sets in the domain to closed sets in the target. This property can alternatively be stated as
\begin{equation} \label{eqn:closedmap}
	\overline{f(C)} \subseteq f(\overline{C}).
\end{equation}
Since the assumed continuity of $f$ also implies the opposite direction of this inclusion, the two sides are actually equal in this case. Next, one proceeds with a proof by contradiction, that is, one assumes that $\partial f(C) \setminus f(\partial C) \neq \varnothing$. Then there must exist a $p \in \mathcal{Z}$ such that
\begin{IEEEeqnarray*}{*l+rl}
	& p \in \partial f(C) \quad {}&{}\wedge \quad p \notin f(\partial C)\\
	~~\Longleftrightarrow & p \in \overline{f(C)} \setminus \mathrm{Int}\pqty{f(C)}  \quad {}&{}\wedge \quad p \notin f(\partial C)\\
	~~\Longleftrightarrow & p \in \overline{f(C)} \quad \wedge \quad p \notin \mathrm{Int}\pqty{f(C)}  \quad {}&{}\wedge \quad p \notin f(\partial C)
\end{IEEEeqnarray*}
where $\mathrm{Int}(A)$ denotes the interior of a set $A$. Since $f$ is a continuous closed map, $\overline{f(C)} = f(\overline{C})$. Therefore,
\begin{IEEEeqnarray*}{*l+rl}
	& p \in \overline{f(C)} \quad \wedge \quad  p \notin f(\partial C) \quad {}&{}\wedge \quad p \notin \mathrm{Int}\pqty{f(C)}\\
	~~\Longleftrightarrow & p \in f(\overline{C}) \quad \wedge \quad  p \notin f(\partial C) \quad {}&{}\wedge \quad p \notin \mathrm{Int}\pqty{f(C)}\\
	~~\Longleftrightarrow & p \in f(\overline{C}) \setminus f(\partial C) \quad {}&{}\wedge \quad p \notin \mathrm{Int}\pqty{f(C)}.
\end{IEEEeqnarray*}
It is always true that $f(A) \setminus f(B) \subseteq f(A\setminus B)$ with equality if and only if $f$ is injective. Thus the statement is still valid if the set on the left-hand side is enlarged
\begin{IEEEeqnarray*}{*r+rl}
	& p \in f(\overline{C}) \setminus f(\partial C) \quad {}&{}\wedge \quad p \notin \mathrm{Int}\pqty{f(C)}\\
	~~\Longrightarrow & p \in f(\overline{C} \setminus \partial C) \quad {}&{}\wedge \quad p \notin \mathrm{Int}\pqty{f(C)}\\
	~~\Longleftrightarrow & p \in f\pqty{\mathrm{Int}\,C} \quad {}&{}\wedge \quad p \notin \mathrm{Int}\pqty{f(C)}
\end{IEEEeqnarray*}
where the last statement creates a contradiction if $f\pqty{\mathrm{Int}\,C} \subseteq \mathrm{Int}\pqty{f(C)}$ holds, which is precisely the definition of an open map $f$. Thus, for continuous maps $f$ which are both open and closed, the relation $\partial f(C) \subseteq f(\partial C)$ must hold.

While this is certainly a valid and illuminating result, having to prove topological openness and closedness every time a new model map is studied can be rather tedious. Hence, it would be advantageous to have a slightly stronger but more practical criterion which model maps can be tested for and from which it already follows that the map is both open and closed.

With this in mind, the target space $\mathcal{Z}$ is now considered in more detail. Once evaluated at a parameter configuration $\theta \in \M$, the model map is a function $y_\model(\openslot;\theta) : \X \longrightarrow \Y$.  Since it would be undesirable in most practical settings for a model map to be unstable with respect to the observation conditions $x \in \X$ in the sense that a small perturbation can result in large and chaotic changes in the predictions of the model, we will restrict our attention to model maps which are continuous with respect to the observation conditions $x \in \X$, i.e.\ the target space $\mathcal{Z}$ is given by 
\begin{equation}
	\mathcal{Z} = C^0(\X,\Y) = \,\Setcond{\vphantom{\big(}y_\model(\openslot;\theta) : \X \longrightarrow \Y}{y_\model(\openslot;\theta)\text{ continuous}}.
\end{equation}
By the definition of global structural identifiability established in \cref{sec:ParameterIdentifiability}, every parameter configuration $\theta \in \M$ must produce a unique prediction $y_\model(\openslot;\theta):\X \longrightarrow \Y$, i.e.\ a model map which is globally structurally identifiable on a set $C \subseteq \M$ is injective on $C$. In addition, by restricting the target of $f$ to $ \N \coloneqq f(C) \subset \mathcal{Z}$, the map trivially becomes surjective onto $\N=f(C)$ such that $f:C \longrightarrow f(C)$ is bijective overall. 

It is well-known that bijective maps are open if and only if they are also closed, that is, if $h : (\M,\mathcal{O}_\M) \longrightarrow (\N,\mathcal{O}_{\N})$ is bijective, one has that
\begin{equation}
	\forall U \in \mathcal{O}_\M : \qquad h(\underbrace{U^c}_{\text{closed}}) = h(\M \setminus U) \overset{\text{injective}}{=} h(\M) \setminus h(U) \overset{\text{surjective}}{=} \underbrace{\N \setminus \overbrace{h(U)}^{\text{open}}}_{\Rightarrow\text{ closed}}
\end{equation}
where $h(U)$ must be open because $U \in \mathcal{O}_\M$ and $h$ is an open map by assumption. Thus, $h$ must be closed and the opposite direction can be shown by a similar argument. Further, it is clear that the openness of a bijective (i.e.\ invertible) map $h$ is equivalent to the requirement that the inverse map $h^{-1}$ is continuous since this means that the preimages of open sets are open. Namely, since $h^{-1}$ exists, one has
\begin{equation}
	\forall U \in \mathcal{O}_\M : \quad h(U) \in \mathcal{O}_{\mathcal{Z}} \qquad \Longleftrightarrow \qquad \forall U \in \mathcal{O}_\M : \quad \preim_{h^{-1}}(U) \in \mathcal{O}_{\N}
\end{equation}
where the right-hand side coincides precisely with the requirement that $h^{-1} : \N \longrightarrow \M$ be continuous.
%
%
%
%
%
Moreover, any space $X$ which can be injectively mapped into a metric space via $h : X \longrightarrow (Y,d_Y)$ can be equipped with the pull-back metric $d_X$ defined by
\begin{equation}
	d_X(x,y) = d_Y\pqty\big{h(x),h(y)}.
\end{equation}
By identifying $X=f(C)$, $Y=C\subseteq \M$ and $f^{-1} = h$, it follows that $f(C) \subseteq \mathcal{Z}$ is metrisable and therefore guaranteed to be Hausdorff. The metric function $d_Y$ that is inherited to $f(C)$ corresponds to the geodesic distance on $\M$, i.e.\ it is computed as the length of the shortest geodesic connecting two given points on $C \subseteq \M$. Finally, the so-called \enquote{closed map lemma} states that continuous maps from compact spaces into Hausdorff spaces are closed and proper \cite{LeeIntroTopManifolds}. The compactness of the set $C \subseteq \M$ can be translated to the requirement that a confidence region $\mathcal{C}_q$ be bounded, which renders its closure $\overline{\mathcal{C}_q}$ compact. This boundedness of a confidence region $\mathcal{C}_q$ further coincides with the property that the model be practically identifiable at the confidence level $q$.




In summary, the mapping of a compact set $C$ under the bijective continuous map $f: C \subseteq \M \longrightarrow f(C)$ renders $f$ closed by the closed map lemma, and simultaneously open via its bijectivity. An appropriate $C \subseteq \M$ is given by the closure of any bounded confidence region $\mathcal{C}_q$ in combination with a model that is globally structurally identifiable on $\overline{\mathcal{C}_q}$, rendering $f$ injective in the first place. Since this means that the identity $\partial f(C) \subseteq f(\partial C)$ holds, it suffices to evaluate the model on $\partial \mathcal{C}_q$ rather than $\overline{\mathcal{C}_q}$ in order to generate the pointwise confidence boundary $\partial \mathcal{B}_q$.

\subsection{Performance and Complexity Details} \label{sec:PerformanceAndComplexityDetails}

The benchmarks in \cref{fig:PerformanceSCP} of \cref{sec:Performance} were performed using \href{https://julialang.org}{\texttt{Julia v1.7.0}}, \href{https://github.com/RafaelArutjunjan/InformationGeometry.jl/releases/tag/v1.10.0}{\texttt{InformationGeometry.jl v1.10.0}} and \href{https://github.com/SciML/OrdinaryDiffEq.jl/releases/tag/v5.68.0}{\texttt{OrdinaryDiffEq.jl v5.68.0}} on an \href{https://www.cpubenchmark.net/cpu.php?cpu=Intel+Core+i5-8265U+\%40+1.60GHz}{\texttt{Intel\,i5-8265U}} mobile processor via single-core computation. 

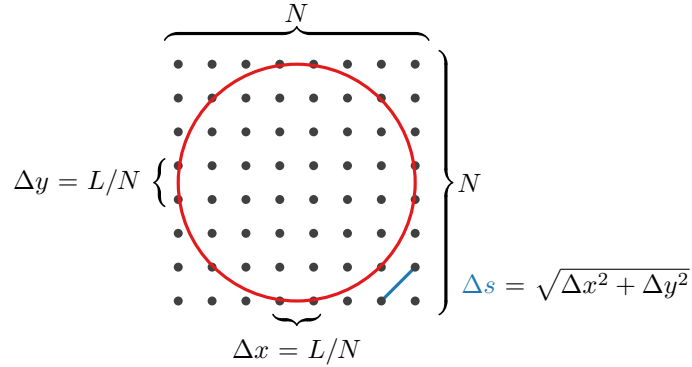
\begin{figure}[!ht]
	\centering
	\PublishGraphic[Saved]{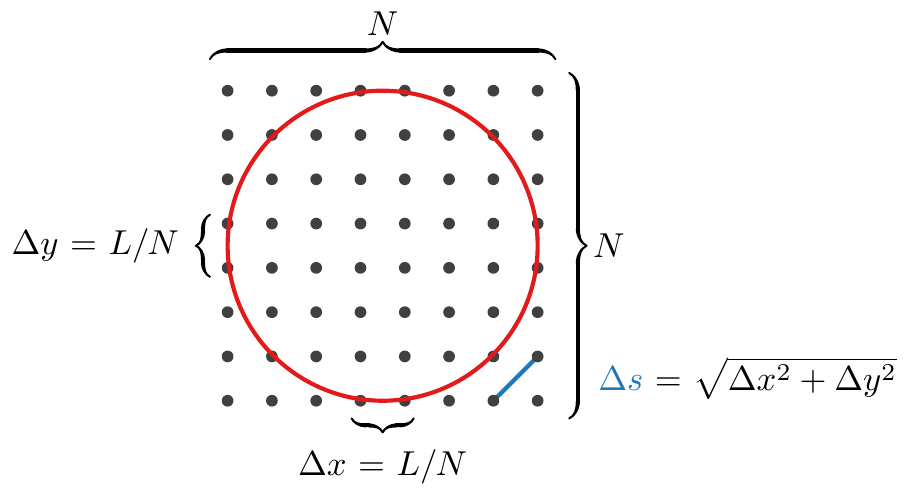}{
		\begin{tikzpicture}[scale=0.45]
			\draw[draw=mycol2, very thick] (7,1) -- (8,2) node [midway, right=2em] {$\textcolor{mycol2}{\Delta s} = \sqrt{\Delta x^2 + \Delta y^2}$};
			\foreach \x in {1,2,...,8}	\foreach \y in {1,2,...,8}
			{\node[mpoint, plotgrey, inner sep=1.2pt] at (\x,\y) {};}
			\draw [decorate, decoration = {calligraphic brace, raise=4pt, amplitude=5pt}, very thick] (8.4,8.4) -- (8.4,0.6) node[midway, right=0.7em] {$N$};
			\draw [decorate, decoration = {calligraphic brace, raise=4pt, amplitude=5pt}, very thick] (0.6,8.4) -- (8.4,8.4) node[midway, above=0.7em] {$N$};
			\draw [decorate, decoration = {calligraphic brace, amplitude=4pt}, very thick] (5.2,0.6) -- (3.8,0.6) node[midway, below=0.5em] {$\Delta x = L/N$};
			\draw [decorate, decoration = {calligraphic brace, amplitude=4pt}, very thick] (0.6, 3.8) -- (0.6, 5.2) node[midway, left=0.5em] {$\Delta y = L/N$};
			\draw[mycol1, very thick] (4.5,4.5) circle (3.5);
	\end{tikzpicture}}
	\caption{Illustration of uniform grid in two dimensions.}
	\label{fig:GridVis}
\end{figure}

The relative tolerance between two points $X$ and $Y$ is usually defined as $\mathrm{rtol} = \norm{Y-X} / \min(\norm{Y},\norm{X})$. In a rough order of magnitude estimation, one can determine the number of uniform grid evaluations required to achieve a given relative tolerance for the location of a given confidence boundary as follows: assume for simplicity that the confidence boundary of interest constitutes a unit circle centered on the origin in coordinates, as depicted in \cref{fig:GridVis}. For linearly parametrised models, this can be achieved exactly using the transformation $f(\theta) = \theta_\text{MLE} + \inv{C} \, \theta$ where $C=\pqty\big{g(\theta_\text{MLE})}^{1/2}$ denotes the Cholesky decomposition of the Fisher metric of the original model evaluated at the MLE such that the modified model is given by $\tilde{y}_\model(x; \theta) = y_\model(x; f(\theta))$. Given this constant radius of one, the denominator of the relative tolerance expression can be simplified to $\min(\norm{Y},\norm{X}) \approx 1$. 
%
%
For a uniform square grid of side length $L$, the maximal distance between any point on the unit circle and its closest neighbouring grid point is given by half of the diagonal distance between grid points, i.e.\ $\mathrm{rtol} \lesssim \Delta s/2 = \sqrt{(\Delta x)^2 + (\Delta y)^2}/2 = L / (\sqrt{2} \, N)$. Assuming, generously, that interpolation of the function values evaluated on the grid points allows one to determine the location of the intermediate crossing to within an accuracy of $\Delta s / 100$, this reduces the necessary number of grid points per dimension to $\mathrm{rtol} \approx \sqrt{2} / (100N)$ for $L=2$. Rearranging, one finds approximately $N \lesssim \num{1.4e-2} \, \mathrm{rtol}$. Choosing for instance $\mathrm{rtol} \overset{!}{=} 10^{-5}$, the total number of required grid point evaluations works out to $N^2 \approx 10^{6}$ in two dimensions.


Although the performance of the grid method can be improved via non-uniform spacings, this does not address the fundamental difference in scaling behaviours as a function of parameter space dimension, i.e.\ $\Complexity{H^{\dim\M}}$ vs $\Complexity{H^{\dim\M-1}}$. Furthermore, the extent of the required sampling grid (i.e.\ a bounding box for the confidence region) is unknown a-priori for non-linearly parametrised models. Whereas the in grid method one evaluates the log-likelihood itself, each \enquote{evaluation} in the integral manifold scheme refers to the gradient of the log-likelihood, i.e.\ the score. However, by using forward-mode automatic differentiation methods, the gradient can be computed very efficiently, which typically requires less than $(\dim \M)$-fold the time of a log-likelihood evaluation while attaining approximately machine precision \cite{ForwardDiffJL}.

After an exact confidence boundary has been computed, it can subsequently be approximated as a polytope consisting of $n$ vertices. By casting this polytope into an appropriate data structure, it is possible to decide whether a given parameter configuration $\theta \in \M$ lies inside or outside the boundary in less than \Complexity{n} time since the log-likelihood does not have to be recomputed \cite{ChazelleCutting}. This allows for extremely performant approximative hypothesis testing which can for instance be used to integrate functions over confidence regions.

	\section*{Author Contributions}

	RA conceptualised the method and formalised the presented approach with frequent inputs from BMS. RA wrote the numerical implementation and created the results presented in this work. The initial idea of investigating likelihood-based confidence regions on parameter manifolds was proposed by BMS. RA wrote a first manuscript draft and all authors revised the manuscript.

	\section*{Acknowledgements}

	This work was funded in part by the Deutsche Forschungsgemeinschaft (DFG, German Research Foundation) under Germany's Excellence Strategy -- EXC-2189 -- Project ID: 390939984. Also, the authors thank Marie Teich, Eileen Giesel, Max Ellinger, Alena Brändle, Tim Litwin and Ricardo Waibel for fruitful discussions as well as constructive criticism regarding the manuscript.

	\nocite{DistributionsJL, DiffEqJL, ForwardDiffJL}

	\bibliographystyle{unsrtnat} 
	\bibliography{Bibliography}

\end{document}